\documentclass[pdflatex,sn-mathphys]{sn-jnl}

\usepackage{lineno}
\usepackage{comment}
\theoremstyle{thmstyleone}%
\theoremstyle{thmstyletwo}%

\theoremstyle{thmstylethree}%

\begin{document}

\title[BR HBC]{Body-Resonance Human Body Communication}


\author[1]{\fnm{Samyadip} \sur{Sarkar}}\email{sarkar46@purdue.edu}

\author[1]{\fnm{Qi} 
\sur{Huang}}\email{huan2065@purdue.edu}

\author[1]{\fnm{Sarthak} 
\sur{Antal}}\email{santal@purdue.edu}

\author[1]{\fnm{Mayukh} \sur{Nath}}\email{nathm@alumni.purdue.edu}

\author*[1]{\fnm{Shreyas} 
\sur{Sen}}\email{shreyas@purdue.edu}

\affil*[1]{\orgdiv{School of Electrical and Computer Engineering}, \orgname{Purdue University}, \orgaddress{\city{West Lafayette}, \postcode{47907}, \state{Indiana}, \country{USA}}}





\abstract{
Seamless interaction between Humans and Artificial Intelligence (AI)-empowered battery-operated miniaturized electronic devices, exponentially transforming the wearable technology industry while forming an \textbf{anthropomorphic artificial nervous system for distributed computing around the human body, demands high-speed low-power connectivity.} If interconnected via traditional radio frequency (RF) based wireless communication techniques, that being radiative, incur substantial absorption losses from the body during non-line-of-sight scenarios and consume higher power ($>$10s of mW). Although being a promising alternative with its non-radiative nature that resulted in  $\sim$100X improvement in energy efficiency ($\sim$sub-10 pJ/bit) and better signal confinement, Electro-Quasistatic Human Body Communication ( EQS HBC) incurs moderate path loss ($\sim$60 to 70 dB), limited bandwidth, and data rate ($\leq$ 20 Mbps), making it less suitable for applications demanding fast connectivity like, high-definition audio-video streaming, human interaction with Augmented-Virtual Reality (AR/VR)-based environments, distributed computing in the network of wearable AI devices. Hence, addressing the \textbf{need for energy-efficient connectivity at rates of 100s of Mbps between wearable devices}, we propose \textbf{Body-Resonance (BR) HBC}, which operates in the \textbf{near-intermediate field region and utilizes the transmission-line-like behavior of the human body channel} to offer \textbf{$\sim$30X} improvement in body channel capacity. Our work sheds new light on the wireless communication system for battery-powered wearables that has the potential to increase the channel gain by \textbf{$\sim$20 dB} with a \textbf{$\sim$10X} improvement in operational bandwidth in comparison to the EQS HBC for reliable digital communication over on-body channels (whole-body coverage area). Experimentally demonstrating BR HBC, we presented \textbf{low-loss ($\sim$40 to 50 dB)} and \textbf{wide-band (hundreds of MHz)} body channels that are \textbf{$\sim$10X less leaky} than antenna-based radiative wireless communication, hence, can revolutionize the design space of new communication systems for a spectrum of applications with wearables in sectors ranging from healthcare, defense, and consumer electronics.
}

\maketitle

\section{Introduction}\label{sec1}
\begin{figure}[ht!]
\centering
\includegraphics[width=\linewidth]{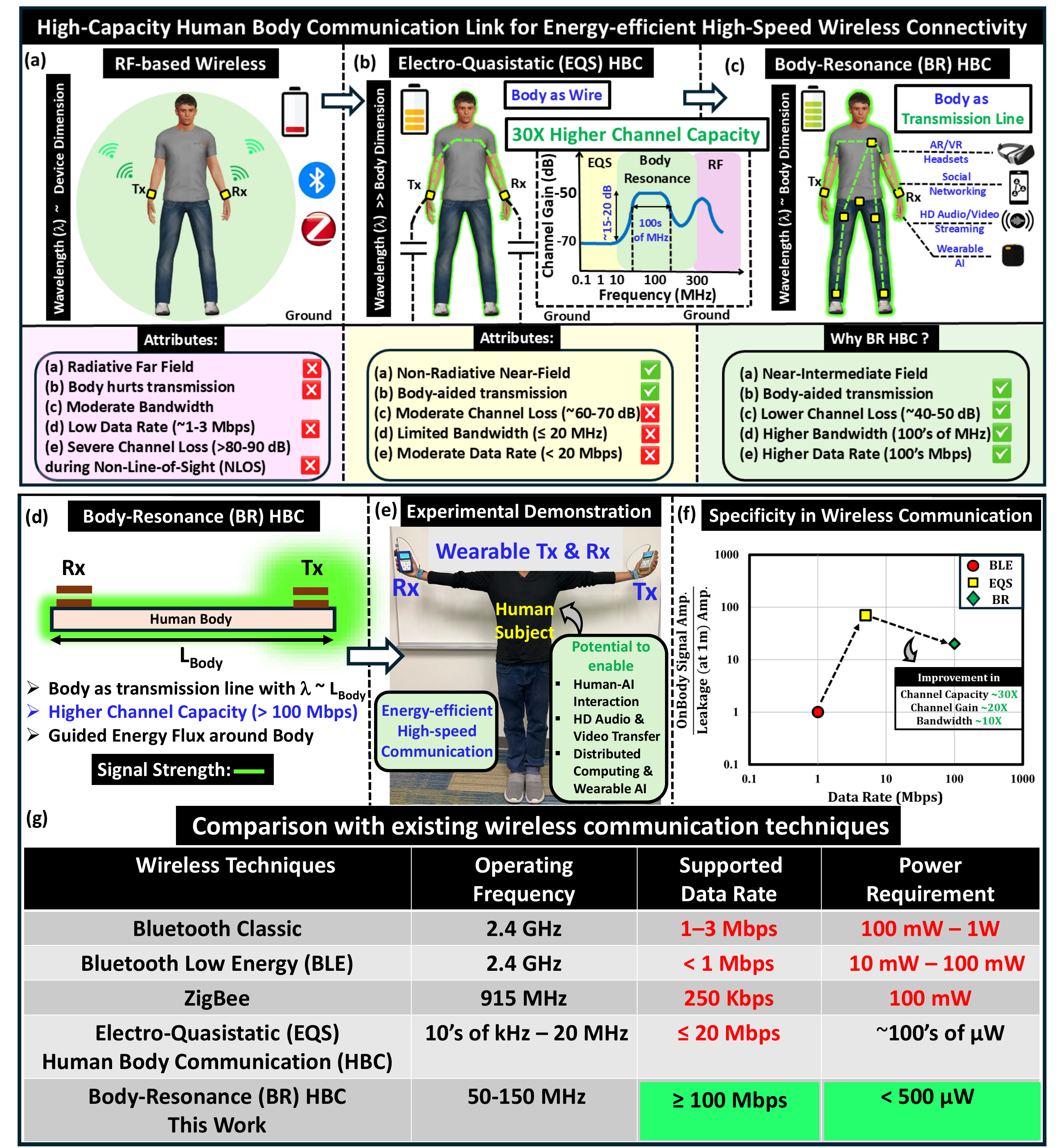}
\caption {\textbf{Need for Energy-efficient High-speed wireless connectivity for wearable devices:} \textbf{a. Traditional radio frequency (RF)-based communication:} radiative with higher power consumption makes it energy in-efficient ($\sim$10 nJ/bit), and besides incurring higher path loss during Non-Line-of-Sight scenarios, it can support a data rate of $\leq$ 3 Mbps. \textbf{b. Electro-Quasistatic Human Body Communication (EQS HBC):} \textbf{Body as Lossy Wire}:  non-radiative, hence energy-efficient ($\sim$sub-10 pJ/bit) though suffers from moderate end-to-end path loss and limited data rate ($\leq$ 20 Mbps) due to constraint on its frequency of operation ($\leq$ 20 MHz) that restricts its bandwidth. \textbf{c. Body-Resonance Human Body Communication (BR HBC):} \textbf{Body as Lossy Transmission Line}: $\sim$15-20 dB higher channel gain for $\sim$10X wider bandwidth makes its Shannon channel capacity $\sim$30X higher compared to EQS. It has potential to support high data rate applications ranging from high-definition audio/video streaming to distributed computing with wearable AI. \textbf{d.} Body as a single-wire transmission line, that couples electromagnetic wave onto a cylindrical conductor.  \textbf{e.} Experimental Demonstration of BR HBC system. \textbf{f.} \textbf{Communication Specificity:} BR HBC offering high data rate while keeping on-body signal order of magnitude higher than off-body leakage. \textbf{g.} Comparison of the BR HBC with the existing techniques for wireless communication. The human figures are made in Open-source software: ‘MakeHuman'.}
\label{fig1}
\end{figure}

The continuous growth of miniaturized computing and communication technologies has revolutionized the cooperation between humans and electronic devices. The increase in the use of such battery-operated smart devices in consumer and medical electronics, remaining a crucial part of the Internet of Things (IoT) \cite{li2015internet, kuzlu2015review,baker2017internet,habibzadeh2019survey}, calls for \textbf{energy-efficient wireless interconnectivity for long-term reliable operation.} Moreover, devices such as smartphones, smartwatches, smart glasses, AI Pocket
Assistants, and AR/VR headsets, while becoming ubiquitous, create a network of wireless devices around our body known as the Internet of Bodies (IoB) \cite{elhayatmy2018internet, sen2020body, chatterjee2023bioelectronic, sen2024human}. With an approximate market share of 10-15$\%$ in the electronics industry, these wearables hold significant potential for developing an \textbf{anthropomorphic artificial nervous system architecture} that could notably transform both the wearable technology and bio-electronics sectors, paving the way for innovative applications and enhanced user experiences. To support high-speed connectivity at low-power yet \textbf{perpetual operation of these devices demand an architectural shift from the state-of-the-art techniques of wireless communication}, illustrated in Fig. \ref{fig1} (a, b, c).
Owing to their radiative nature, the traditional radio frequency (RF)-based wireless communication methods among these devices consume higher power ($>$10s of mW) and experience severe transmission losses ($\sim$80-90 dB or even higher) from the surroundings and absorption from the body \cite{garn1995present, rehman2010investigation} in non-line-of-sight scenarios, making them less suitable for IoB applications that demand superior energy efficiency and physical security. Besides, their limited maximum data rates ($\leq$ 3 Mbps) restrict their scope of application. 

Addressing the need for energy-efficient connectivity, Human Body Communication (HBC) leverages the electrically conducting properties of the human body to facilitate energy-efficient information exchange among wearable devices. 
The exploration of different modalities of HBC over the last couple of decades has unveiled enabling data transfer via electric fields \cite{song2013review, callejon2015measurement, maity2018bio, maity2019bodywire, li2021body}, magnetic fields \cite{park2019sub, wen2021channel, nath2022understanding}, and acoustics \cite{zhang2017bioacoustics, bos2018enabling, gerguis2021enabling}.
\subsection*{Cons of existing HBC modes: Need for new modality}
Being transparent to the quasistatic magnetic field (frequency $<$ $\sim$30 MHz), the presence of the human body doesn't influence the channel behavior when the signal transmission occurs via the means of the magnetic field. Moreover, the \textbf{inductive links}, being susceptible to the relative separation and the orientation mismatch between the transmitter and receiver coils, cause sharper attenuation in transmission channel gain that makes magneto-quasistatic HBC \textbf{not an optimal choice for long-range on-body channels}. Furthermore, the body tissues, being conductive, attenuate the propagating high-frequency Magnetic HBC fields because the induced eddy currents hinder the efficient propagation of waveguide modes within the body, making it not a preferred choice for high-speed body-communication links. Moreover, the limited coverage area and inability to support higher data rate communication restricts the application scope for transmitting information via acoustic signals through the human body.

Among the different modes of electric HBC, the galvanic HBC has limited coverage due to the dipole-dipole interaction governing its operation, which reduces the channel gain away from the transmitter \cite{modak2022bio}. On the contrary, the capacitive HBC operating in the Electro-Quasistatic (EQS) frequency regime emerged as a promising choice for its frequency-independent approximately consistent channel gain across the entire body channel with high-impedance capacitive termination at the receiver \cite{maity2018bio}. This makes capacitive EQS HBC suitable for on-body communication links with higher link margins. In EQS capacitive HBC, the signal gets coupled into the body from a transmitter using single-ended excitation comprising of a signal electrode that remains in contact with the body, and the ground electrode of the device is left floating. With its lower carrier frequency (typically $\leq$ 20 MHz) resulting in the operating wavelength ($\geq$ 15 m) being order of magnitude larger than human body dimensions ($\leq$ 2 m), in capacitive EQS HBC, the human body, remaining at the same potential throughout, constitutes the forward path and the parasitic couplings from the ground of the communicating devices to the earth ground form the return path \cite{nath2019toward}.
Its non-radiative nature, which results in better signal confinement, makes this mode energy efficient ($\sim$ sub-10 pJ/bit) and physically secure over RF. However, its limited channel capacity ($C = B \log_2 (1+\frac{S}{N})$, according to Shannon Hartley Theorem) due to its lower operating frequency ($\leq$20 MHz) and moderate channel loss ($\sim$60 to 70 dB) becomes a bottleneck as it falls short in supporting high data rate applications requiring 100s of Mbps operational data rate. Addressing these limitations, this paper introduces a new modality for body communication, where the human body acts as a lossy transmission-line-like channel that utilizes the electromagnetic resonant phenomena to offer a $\sim$30X improvement in the body channel capacity. A comparative analysis of the proposed BR-HBC, with state-of-the-art techniques for wireless communication, is presented in Fig. \ref{fig1} (g). This approach enables low-loss ($\sim$40 to 50 dB), high-speed (100s of megabits per sec) connectivity for wearable devices that inspire the emergence of various novel body-centric technologies that include but are not limited to high-definition audio-video streaming, distributed computing with wearable AI, and seamless interaction between human and augmented-virtual reality-based devices.

\section{Background $\&$ Related Works}
Previously, very few works attempted to enable high-frequency communication modes around the human body for energy-efficient high-speed connectivity among wireless body-area network devices. 
Considering the human body as a monopole antenna, Kibret et al. \cite{kibret2014human} studied its antenna properties and its effect on human body communication over a frequency range of 1 MHz to 200 MHz. However, their use of a battery-powered vector network analyzer (VNA) and baluns for isolating the ground of the transmitter from the receiver does not emulate the body channel measurements in the wearable-to-wearable scenario . Using a similar approach, Li et al.\cite{li2017evaluation} investigated wireless inter-human signal transmission for frequencies between 1 MHz and 90 MHz. Nevertheless, their use of wall-connected VNA does not take the parasitic coupling between the floating ground and the earth’s ground (i.e., the return path capacitances of Tx and Rx) into consideration. Consequently, these approaches do not replicate measurements with wearable devices, thereby limiting the direct applicability of the insights gained from these studies. Subsequently, Park et al. \cite{park2016channel} presented path loss measurements for capacitive HBC using miniaturized battery-powered wearables and a moderate impedance matching network to maximize power transfer over a frequency range from 20 to 150 MHz. In their study on Body Channel Communication, Bae et al. \cite{bae2012signal} investigated how signals travel on the surface of the human body. They used a dipole model-based analysis to study components of the electric field, such as the near-field quasi-static coupling, reactive radiation, and the surface wave far-field across frequencies from 100 kHz to 100 MHz and distances of up to 1.3 m on the body. However, their methods, which involved using a wall-connected spectrum analyzer and balun, led to an optimistic estimation of the path loss and shifted the resonance peak of the human body to a lower frequency range, between 30 and 50 MHz. The experimental characterization of the body channel for capacitive HBC and its dependency on termination impedance using miniaturized wearable devices over a broad frequency range (100 kHz to 1 GHz), was done by Avlani et al. \cite{avlani2020100khz}. Their utilization of miniaturized devices with the floating ground though presents realistic path loss estimations, but the scope was limited as it does not analyze the factors influencing the channel variability. Afterward, while analyzing security and interference aspects of capacitive inter-body communication, a conceptual understanding of the body channel characteristics via identifying three regions, namely EQS, Body-Resonance, and Device Resonance, was proposed by Nath et al. \cite{nath2021inter} through wide-band measurements up to 1 GHz using wearable devices. They showed that by increasing the frequency beyond the EQS regime, the human body starts exhibiting dimensional resonance between frequencies ranging from 60 MHz to 150 MHz when the operating wavelength starts approaching the dimension of the human body. However, finding the exact location of the peaks and their variability,  along with the factors that influence their appearance in the channel transfer characteristics, was beyond the scope of this work and requires a detailed investigation. Moving consequently higher in frequencies, such as in the range beyond 1 GHz, they observed device resonance that refers to the regime where the operating frequency is high enough to make the wavelength comparable to the size of the devices i.e., when capacitive HBC starts resembling RF-based communication with electrodes of the devices becoming antennas. Recently, while focusing on the effect of material properties on the body-resonance phenomena, Sarkar et al. provided wave-impedance based understandings \cite{sarkar2024wearable} and demonstrated it \cite{sarkar2024material}. However, there is no fundamental study on the body-resonance frequency regime in the literature that would improve our understanding of the phenomenon itself and potentially guide the design of energy-efficient transceivers for high-speed connectivity. Though, due to the near-intermediate field of operation, increased radiative component from the transmitter and humans, the body channel in the BR frequency range appears to be more leaky than the EQS mode of communication, the crucial need to enhance the channel capacity has prompted us to ask: \textbf{Can we utilize the electromagnetic waves, resonant to human body's dimensions to facilitate energy-efficient, high-speed connectivity at speeds of hundreds of megabits per second among body area network (BAN) devices?}

In this work, we demonstrate BR HBC that enables energy-efficient fast connectivity for wearables by utilizing the transmission line-like behavior of the human body, and the contributions of the proposed modality are summarized below:
\newline\textbf{Technical advancement through BR-HBC:} 
\newline\textbf{(i) Enabling Higher Channel Capacity:} This work enables high-capacity body channel (\textbf{$\sim$30X} higher than EQS) that offers low-loss, wide-band high-speed (\textbf{$\sim$100s of Mbps}) wireless connectivity that will potentially inspire the design of energy-efficient transceivers and emergence of novel body-centric applications using wearables that facilitate augmented living.
\newline\textbf{(ii) Body as a Lossy Transmission Line:}
Presenting a new perspective by conceptualizing the human body as an unbalanced lossy transmission line, we consider the body acts as a signal conductor, while the earth's ground serves as an additional conductor. These understandings are supported by numerical electromagnetic simulations and experiments, demonstrating the electrically distributed characteristics of the body channel in the BR frequency range.
\newline\textbf{(iii) Body-channel characterization for reliable communication:}
With the body being a lossy transmission line, the variability in the body-channel characteristic in the BR frequency regime depends on several factors, influencing the transmission line parameters that remain decisive in optimizing the channel capacity. The body channel is characterized to estimate the capacity of various body-communication links based on the receiver sensitivity and the adequate data rate for reliable communication.
\newline\textbf{(iv) Leakage profile analysis of the proposed body-communication link:} Specificity in information exchange remains crucial while analyzing a wireless communication channel. This work presents the results from the leakage measurements around an on-body communication link and provides a comparative analysis of its signal confinement behavior with traditional radiative communication like Bluetooth.

In the following section, we delve into the influence of the human body on electromagnetic signal propagation behavior, illustrated in Fig. \ref{fig2}.
\section{Signal Propagation in EQS and above}
\begin{figure}[ht!]
\centering
\includegraphics[width=\linewidth]{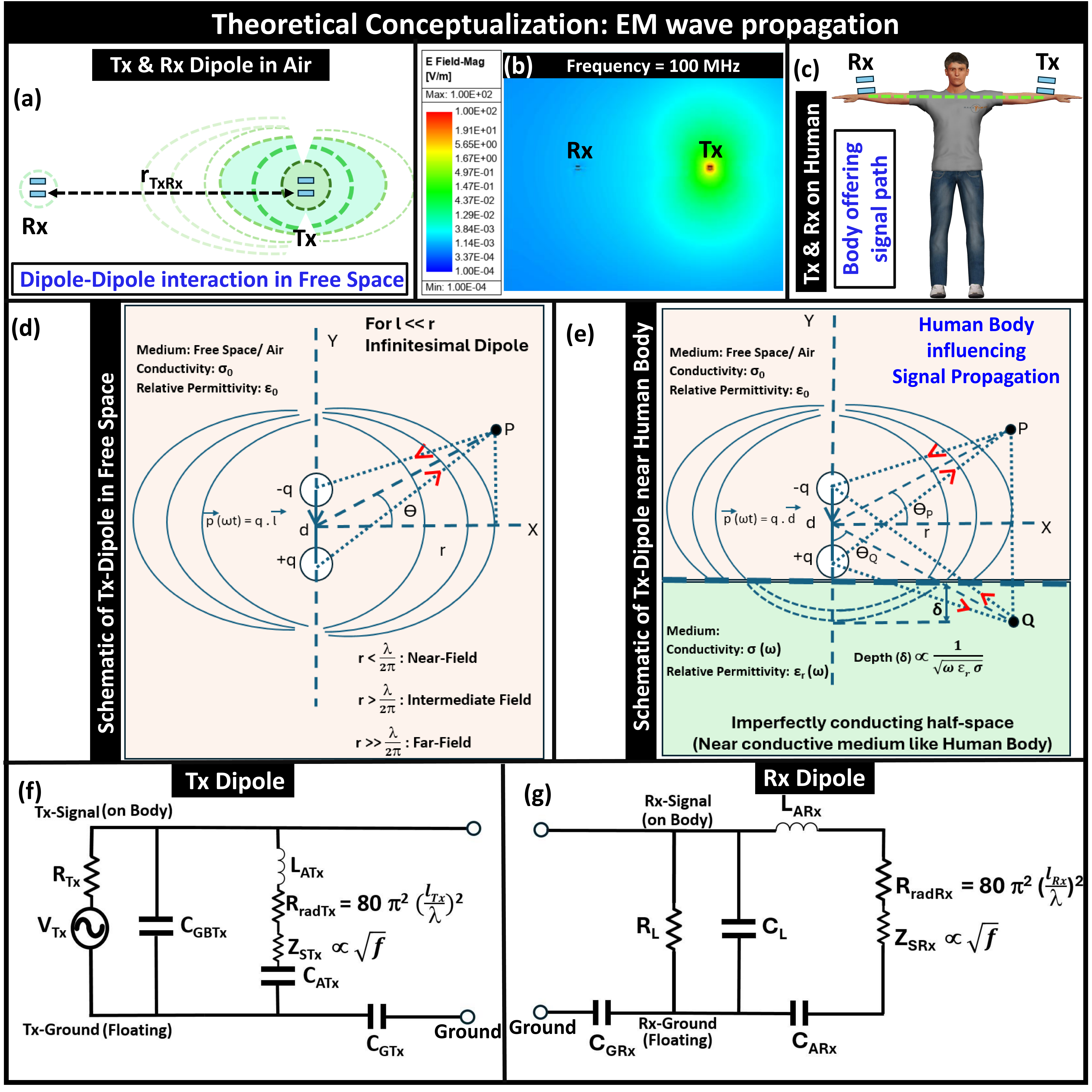}
\caption {\textbf{Conceptualizing EM wave propagation behavior in the presence of the Human Body:} \textbf{a.} Tx and Rx dipole communicating in Air. \textbf{b.} Electric-field plot of the Tx-Rx interaction at 100 MHz. \textbf{c.} Human Body Communication with Tx and Rx on the body. Schematic representation of the radiation from Tx-Dipole: \textbf{d.} in Free-space /Air, \textbf{e.} in the presence of the human body. \textbf{Conceptual Circuit model:} \textbf{f.} Tx-dipole, \textbf{g.} Rx-dipole.}
\label{fig2}
\end{figure}

 Aiming to simplify our understanding of the complex nature of EM wave propagation around the body, we start from the radiation theory of a Hertzian dipole while considering our wearable transmitter and receiver as an infinitesimal dipole (under the following assumptions: the size of the transmitter i.e., effective dipole length (l) $<<$ operating wavelength ($\lambda$)) communicating in the air (i.e., in an infinite space). Now,  depending on the ratio of the field observation point (r) to $\lambda$, the propagation region around the Tx dipole can be sub-divided under the following categories: (a) near-field ($\beta r << 1$), (b) intermediate field ($\beta r > 1$), and (c) far-field ($\beta r >> 1$), where $\beta$ represents the wavevector as schematically presented in Fig. \ref{fig2} (d). 
\newline With $l$ representing the thickness of the transmitting device (i.e., the separation between the signal and the ground plate) and I denoting the current supplied to the signal electrode, $\eta$ standing for medium's wave impedance, ($r, \theta, \phi$) standing for the spatial location of the point of observation, the following well-known equations depict the electromagnetic fields from a Hertzian dipole: 
\begin{subequations}
\begin{equation}
\centering
    E_r = \eta\frac{I_0 l \cos \theta}{2 \pi}\left[\frac{1}{r^2} + \frac{1}{j \beta r^3}\right] e^{-j \beta r}\label{first:a}
\end{equation}
\begin{equation}
\centering
E_\theta = j \eta \frac{\beta I_0 l \sin \theta}{4 \pi} \left[\frac{1}{r} + \frac{1}{j \beta r^2} - \frac{1}{(\beta^2r^3)}\right] e^{-j \beta r}\label{first:b}
\end{equation}
\begin{equation}
\centering
E_\phi = 0 \label{first:c}
\end{equation}
\begin{equation}
\centering
H_r = H_\theta = 0 \label{first:d}
\end{equation}
\begin{equation}
\centering
H_\phi = j \frac{\beta I_0 l \sin \theta}{4 \pi} \left[\frac{1}{r} + \frac{1}{j \beta r^2}\right] e^{-j \beta r}\label{first:e}
\end{equation}
\end{subequations}

In the near-field region since $\beta r << 1$, the term involving $\frac{1}{r^3}$ becomes dominant. Therefore, at lower frequencies, i.e., in Electro-Quasistatics (EQS) (10s of MHz range or less), where the wavelength of the signals is significantly larger than the dimensions of the communicating devices and the human body, the dominant mode of transmission is the quasistatic near-field. Hence, the transmitter can be simplified as an electro-quasistatic dipole exciting the human, which can be approximated as a node as the consistent potential exists throughout the body. Besides, below a certain frequency ($f_c$), the voltage developed by the magnetic fields ($H_{\phi}$), associated with an electric dipole, is usually ignored as there are no closed current loops (i.e., no formation of the magnetic dipole) present at the electrodes of the devices. Now, when the Tx and Rx dipoles are mounted on the human body, the EQS signal from the Tx gets capacitively coupled to the human body and picked up at the Rx. This enables the EQS HBC, which has been broadly explored \cite{maity2018bio, nath2019toward, datta2021advanced}. 
However, as the operating frequency increases beyond the EQS range, the near-field quasistatic approximation of Tx-dipole no longer holds accurately as its reactive part of the transmitted energy starts influencing signal transmission. Therefore, the influence of the intermediate field also need to be emphasized. The boundary between the near and the intermediate field can be conceptualized via a distance (r$_d$) from the source, namely the radiation zone which can be calculated as, r$_d$ = $\frac{1}{\beta}$ = $\frac{2\pi}{\lambda}$. While, in the intermediate field region, the term with $\frac{1}{r^2}$ predominates, in the far field, the term with $\frac{1}{r}$ decides the field decay characteristics. The interaction between electric dipoles can be approximately captured in conceptual circuit models, shown in Fig. \ref{fig2} (f, g), where the inductances (L$_{ATx}$, L$_{ARx}$), capacitances (C$_{ATx}$, C$_{ARx}$) and radiation resistance (R$_{radTx}$, R$_{radRx}$ $\propto$ $\frac{1}{\lambda^2}$) are used model the radiative behavior of the Tx and Rx. This model gets reduced to the simplified model of an EQS capacitive Tx and Rx at lower frequencies (i.e., f $\leq$ 20 MHz).

In the BR frequency regime, assumed to be lying in the range of 30 MHz to 300 MHz, where the wavelength (10 cm $<$ $\lambda$ $<$ 1 m) is comparable to the dimensions of the human body ($\lambda$ $\leq$ 2 m), the transmission behavior can be considered to be dominated mostly by near and intermediate fields and sometimes far-field. Let's consider a practical scenario of a wearable transmitter positioned on the surface of the human body, with its signal electrode in contact with the skin and the ground electrode left floating. This transmitter setup can be approximated as a vertical electric dipole (VED) with its orientation perpendicular to the axis of the single-cylinder approximation of the human model and is situated at the body surface that behaves like an imperfectly conducting half-space, shown in Fig. \ref{fig2} (e). The value of quasistatic fields at points not very close to the VED but at distances much less than the free space wavelength ($\lambda_0$) can be approximated in the Lorenz gauge to appear at the analytical solutions of quasistatic scalar electric potential from the expression of the magnetic vector potential, provided in the work of Kibret et al. Owing to the dependency of the induced electric field on the scalar and vector potential, the choice of gauge becomes crucial in estimating the analytical solution. To verify their hypothesis, they used a battery-powered vector network analyzer (VNA) and baluns to isolate the ground of communicating devices. Previous studies indicate that the use of baluns does not replicate the real-world wearable-to-wearable measurements, as it does not take the return path capacitances of Tx and Rx into account. Therefore, it is essential to analyze with proper ground isolation between the transmitter and receiver.

\begin{figure}[ht!]
\centering
\includegraphics[width=\linewidth]{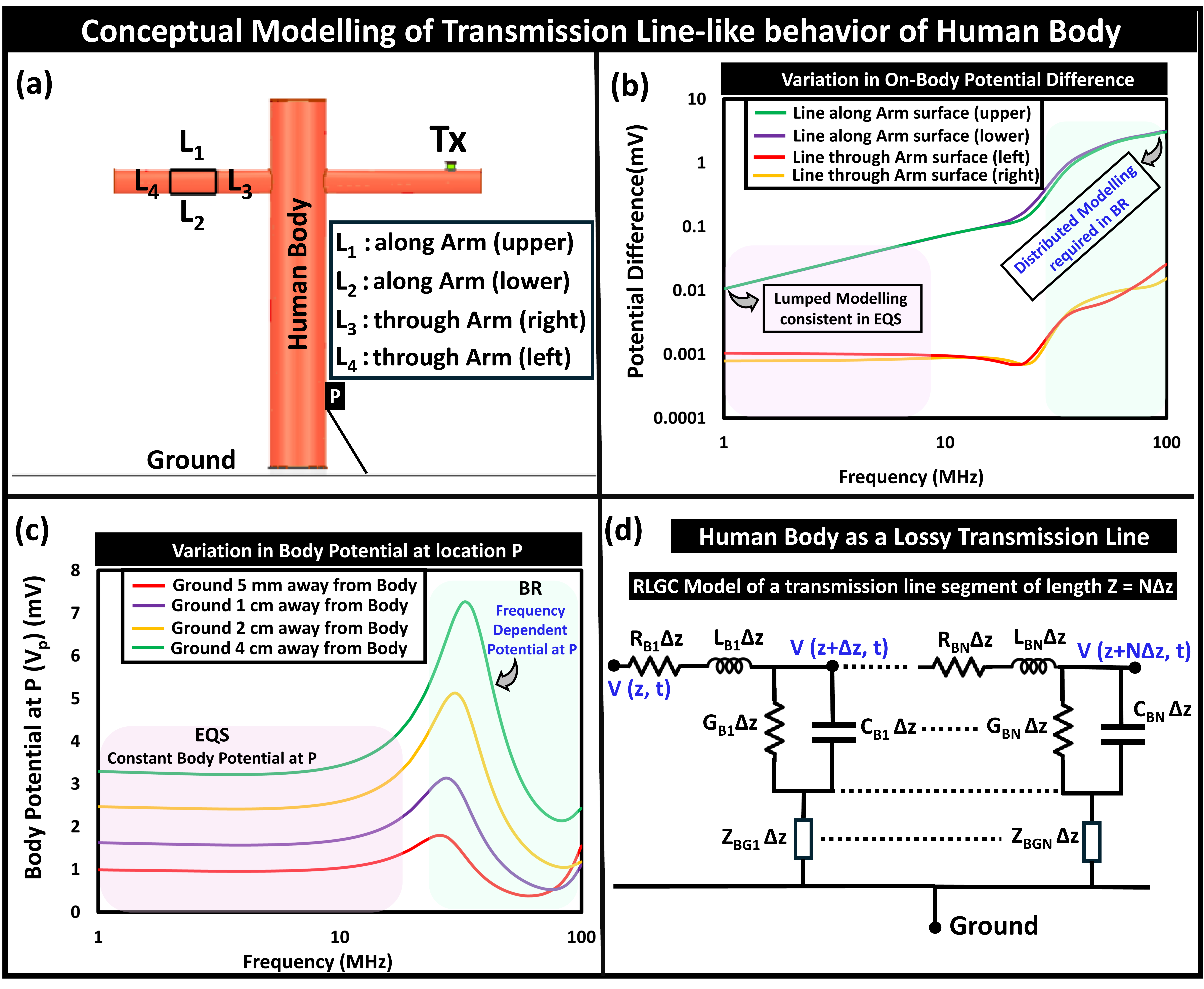}
\caption {\textbf{Conceptual Modelling of Human Body as a Lossy Transmission Line} \textbf{a.} Numerical simulation setup for electromagnetic analysis. \textbf{b.} Variation in the potential difference between two locations, highlighting the need for distributed modeling of the body in BR. \textbf{c.} Variation in Body potential at location P relative to the ground, illustrating frequency dependency of potential P.  \textbf{d.} Conceptual RLGC circuit of the human body, modeled as a lossy unbalanced transmission line with the human body being the signal conductor and the earth's ground as the ground conductor. 
}
\label{fig12}
\end{figure}

With the dominance of the parasitic return path on the estimated path loss with wearable devices, previous studies on capacitive EQS HBC analyzing simplified lumped elements-based biophysical models provided fairly consistent results for the body channel characteristic. However, the increased complexity of the propagation mechanism in the BR frequency regime demands more of a distributed modeling of the human body which can be conceptualized in a unbalanced \textbf{lossy transmission line-like} behavior with the body acting as a signal conductor and the earth's ground acting as a ground conductor, illustrated in Fig. \ref{fig12}. The increased variation in the potential difference between two locations on the human body with an increase in frequency highlights the requirement for distributed modeling of the body channel in the BR, which can otherwise be regarded as a lumped node in the EQS regime, depicted in Fig. \ref{fig12} (b). Moreover, the frequency-dependent potential variability at a point P in BR, delineated in Fig. \ref{fig12} (c), presents the influence of the ground location. A conceptual model for the body channel behavior as a lossy transmission line is presented in Fig. \ref{fig12} (d). In very few previous attempts to capture the dispersion and attenuation characteristics of the body channel, a transmission line model (TLM) using infinite structures and a periodic transmission line Model of finite length (limited only over a frequency range from 1 kHz to 1 MHz) \cite{lodi2020periodic}, were studied. Although developing a detailed biophysical model capturing the body channel variability in the BR regime is beyond the scope of this paper, serves as a motivation for future work. The \textbf{field around the transmission line-like human body}, being a function of the total axial current, changes based on the followings: \textbf{ (i) incident field, (ii) dielectric properties of the body tissues} (relative permittivity ($\epsilon_r$) and conductivity ($\sigma$)), \textbf{(iii) dimensions of the human model (length, width and thickness of the transmission line)} (i.e., positional and frequency dependent variability in $R_{Bi}$, $L_{Bi}$, $G_{Bi}$, and $C_{Bi}$ where i = 1, 2, ...N, in Fig. \ref{fig12} (d)), and \textbf{(iv) impedance (Z$_{BG}$) from the subject's body to the earth's ground}. Hence, analyzing the influence of these factors is crucial for understanding the body channel behavior.

\section{Enabling Energy-efficient High-speed Connectivity via BR HBC}
To investigate the influence of the aforesaid factors on channel capacity, we take a reductionist approach. This section discusses the potential for using the human body as a high-capacity channel for wireless communication between transmitters (Tx) and receivers (Rx). Specifically, we analyze a single cylindrical model that is 180 cm long, equivalent to the height of a human, before progressing to a full-body model as follows:

 \begin{figure}[ht!]
\centering
\includegraphics[width=\linewidth]{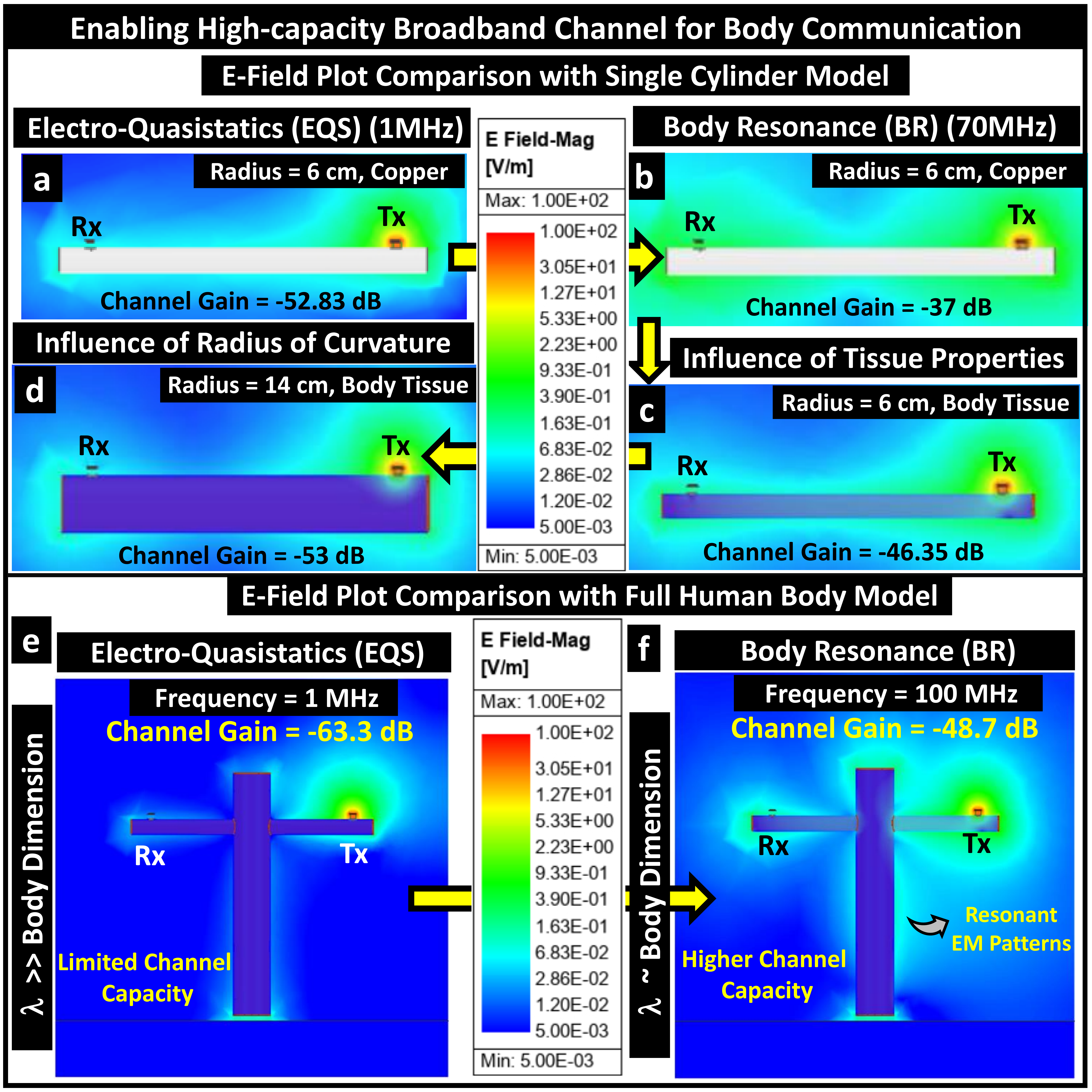}
\caption {\textbf{Reductionist approach to analyze electromagnetic resonance of conducting medium:} \textbf{Single-cylinder model:} Tx $\&$ Rx on a copper cylinder, emulating human arm: \textbf{a.} in EQS regime, \textbf{b.} in BR regime. \textbf{Dependency on Tissue Properties:} \textbf{c.} Cylinder material replaced with body tissue (skin and muscle). \textbf{Dependency on Limb Thickness:} \textbf{d.} The cylinder radius increased to emulate the human torso. \textbf{Cross-cylindrical Human Body Model:} \textbf{E-Field Plot comparison:} \textbf{e.} Lower field strength at Rx indicates limited channel capacity. \textbf{f.} Higher field strength at Rx denotes higher channel gain that results in an improvement in channel capacity.}
\label{fig3}
\end{figure}

\begin{itemize}
    \item We began by considering a scenario where the Tx and Rx, separated by a distance of 150 cm, are mounted on a copper cylinder with a 6 cm radius, similar to the dimensions of a human arm. After numerically simulating the single-cylinder model for electromagnetic analysis in ANSYS HFSS at a frequency of 1 MHz, the simulated channel gain is determined to be -52.83 dB. The electric field plot, depicted in Fig. \ref{fig3} (a), illustrates the confinement of the transmitted signal around the cylinder.
    \item Next, the operating frequency is increased to 70 MHz, i.e., to the peak frequency in the transfer characteristic, to observe the effect of the resonance phenomena of the copper cylinder. We notice a $\sim$15 dB improvement in the simulated channel gain as it becomes -37 dB. The field plot, shown in Fig. \ref{fig3} (b), depicts the improvement in E-field strength at the Tx and Rx and the formation of a E-field null in between.
    \item Subsequently, to investigate the influence of tissue properties on the resonant behavior of the cylindrical resonator, we change the material property of the cylinder to the dielectric properties (conductivity ($\sigma$) and relative permittivity ($\epsilon_r$)) of the body tissues (skin and muscle), adopted from the works of Gabriel et al. \cite{Gabriel_1996}. We observe a $\sim$9 dB attenuation in channel gain due to a combination of the following: (i) The higher relative permittivity ($\epsilon_r$) of the tissues than air that reduces the propagation velocity (v$_p$ $\propto$ $\frac{1}{\sqrt{\epsilon_r}}$) and  (ii) the conductivity ($\sigma$) of the body tissues that is several orders of magnitude lower than copper. Besides, the E-field plot, presented in Fig. \ref{fig3} (c), shows the resonance of the tissue cylinder with a reduced quality factor (Q) from the dielectric loss. The reduced Q of the human body, being an imperfect resonator, gives rise to a relatively higher loss wide band channel (Bandwidth (BW) = $\frac{f_c}{Q}$) than a conducting cylinder while reducing its radiation efficiency.
    \item Furthermore, to study the effect of change in limb thickness (i.e., thickness of transmission line), in other words, the radius of curvature of the body cylinder, we increase the cylinder radius to 14 cm to mimic the lateral dimension of the human torso. An attenuation in channel gain is observable as it goes down to -53 dB due to higher path loss from the increased tissue thickness. The field plot highlights the reduced magnitude of the E-field strength at the Rx in Fig. \ref{fig3} (d).
    \item  Finally, a comparative analysis of the field plot with the entire cross-cylindrical human body model between EQS (at 1 MHz) and BR (at 100 MHz) is provided in Fig. \ref{fig3} (e, f). Though offering better signal confinement with its non-radiative behavior, EQS HBC limits the application's data rate with its lower bandwidth. The increased E-field strength around the body highlights the benefits of enabling HBC in the BR frequency regime as it supports a high-capacity channel. 
\end{itemize}


\section{Results}\label{sec2}
The following section depicts the results from numerical simulations and measurements with wearable devices that illustrate the benefits of enabling human body communication in the Body-Resonance frequency regime. 
\subsection*{Numerical Simulation Results}
To support the developed understanding from the theory, as described above, here we present the results of Electromagnetic (EM) simulations performed within a Finite Element Method (FEM) based electromagnetic solver, namely High-Frequency Structure Simulator (HFSS) from ANSYS. Simulations are executed on a simplified cross-cylindrical human body model, whose accuracy has been confirmed previously through comparison of its field and current density distribution with a detailed model by Maity et al \cite{maity2020safety}. 
With an input excitation as an ideal AC voltage source of amplitude (V$_{in}$) 1 V, received voltage (V$_{Rx}$) (in dBV) variations are calculated from the induced electric field.

  \begin{figure}[ht!]
\centering
\includegraphics[width=\linewidth]{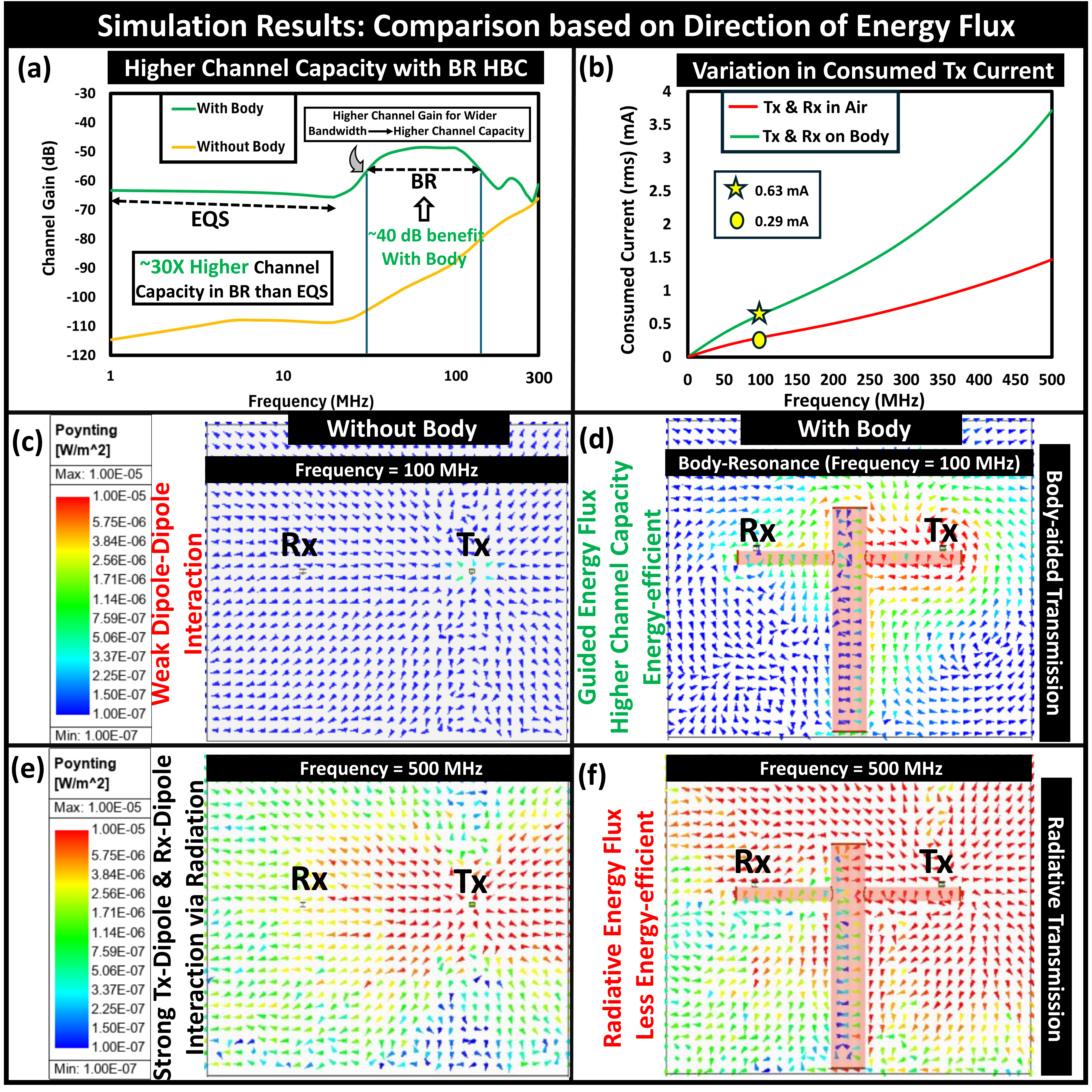}
\caption {\textbf{Human body as a energy-efficient high-speed communication channel in BR:} \textbf{a.} Human body providing $\sim$10X more bandwidth and $\sim$20 dB higher channel gain that leads to $\sim$30X improvement in body channel capacity over EQS regime. \textbf{b.} Transmitter current variation depicting the need for careful choice for operational frequency range that optimizes the trade-off between energy-efficiencies and faster connectivity.
\textbf{c.} In the absence of humans, the Tx and Rx dipoles weakly interact via dipole-dipole interaction. \textbf{d.} Human body supporting guided energy flux density while improving channel capacity in BR. \textbf{e.} Stronger Tx-Rx interaction exists via radiation at a higher frequency. \textbf{f.} Body acting as an extended ground boosts the radiation from Tx, resulting in non-guided energy flux density.
}
\label{fig4}
\end{figure}

\subsubsection*{Human Body influencing Signal Transmission}
In the absence of the human body, when the devices (Tx $\&$ Rx) wirelessly communicate in air, the dipole-dipole interaction between Tx and Rx decides the received signal strength, i.e., the coupled field strength in EQS and radiated field strength when beyond EQS (in the EM frequency regime). However, with its electrically conducting nature, the presence of the human body offers a conducting path for signal transmission during the on-body positioning of the Tx and Rx and decisively influences the received signal strength. The body channel characteristic, presented in Fig. \ref{fig4} (a), depicts the benefits of enabling BR HBC as it offers $\sim$40 dB higher channel gain compared to device-to-device coupling (i.e., when the devices couple through the air). Besides offering higher channel gain, being a conductor of conductivity ($\sigma$ $\sim$0.6-0.7 S/m) several orders of magnitude smaller than metals, the human body supports wider operational bandwidth (due to its lower quality factor resulting from the Body-Resonance), that enables high channel capacity. The location of the peak in the channel characteristic between $\sim$50-150 MHz confirms its appearance from the resonance phenomena of the human body since the devices are small enough (i.e., $\leq$ 3 cm) to introduce peaks from their antenna characteristics (i.e., at 2.5 GHz if quarter-wave dipole and 5 GHz if half-wave dipole in air) in the channel behavior in BR frequency regime. However, the variation in the consumed transmitter current, portrayed in Fig. \ref{fig4} (b), depicts that in an attempt to enable a high data rate, the human body starts drawing more current from the transmitter in the BR regime, which may raise some concerns from the user's safety viewpoint which demand an analysis as per the guidelines issued by the International Commission on Non-Ionizing Radiation Protection (ICNIRP) \cite{ICNIRP2020}, and hence addressed in a subsequent subsection.
Moreover, the benefits of BR HBC can be visualized via observing the direction of energy flux density, i.e., the Poynting vector plot (\textbf{$S = E \times H^*$}), highlighting strong guided energy flux density along the body surface, i.e., directive signal flow that strengthens the field at Rx at BR, illustrated in Fig \ref{fig4} (d). However, with a further increase in the operating frequency, the dipole-dipole interaction between Tx $\&$ Rx in the air starts becoming stronger, shown in Fig. \ref{fig4} (e), which gets boosted even further by the presence of the human body, making the energy flux density more non-guided, illustrated in Fig. \ref{fig4} (f). This happens due to a combination of the following: (i) a shift in the device peaks to a lower frequency due to the higher dielectric index of the surrounding body tissues (i.e., f$_c$ $\propto$ $\frac{1}{\sqrt{\epsilon_{tissue}}}$), that occurs due to an increase in the effective electrical length of the communicating devices (L/$\lambda_{eff}$), (ii) the human body as a large conductor acting as an extended ground to boost the gain of the transmitting and receiving antennas, and (iii) some part of the wave gets guided along the body surface as a surface wave. Though offering higher channel gain, the highly non-guided energy flux density, depicted in Fig. \ref{fig4} (f), makes it radiative transmission and not an energy-efficient choice for the long-term operation of battery-powered wearables. Animation plots, capturing the variation in E-field and Poynting Vector (Energy Flux Density) in frequency domain are included with this manuscript as Supplementary Movie 1 and its link is provided here: \url{https://github.com/SparcLab/BodyResonanceHBC}. The transient variation in the E-field at a selected BR frequency of 100 MHz is captured in Supplementary Movie 3 provided in the above link.

\begin{figure}[ht!]
\centering
\includegraphics[width=\linewidth]{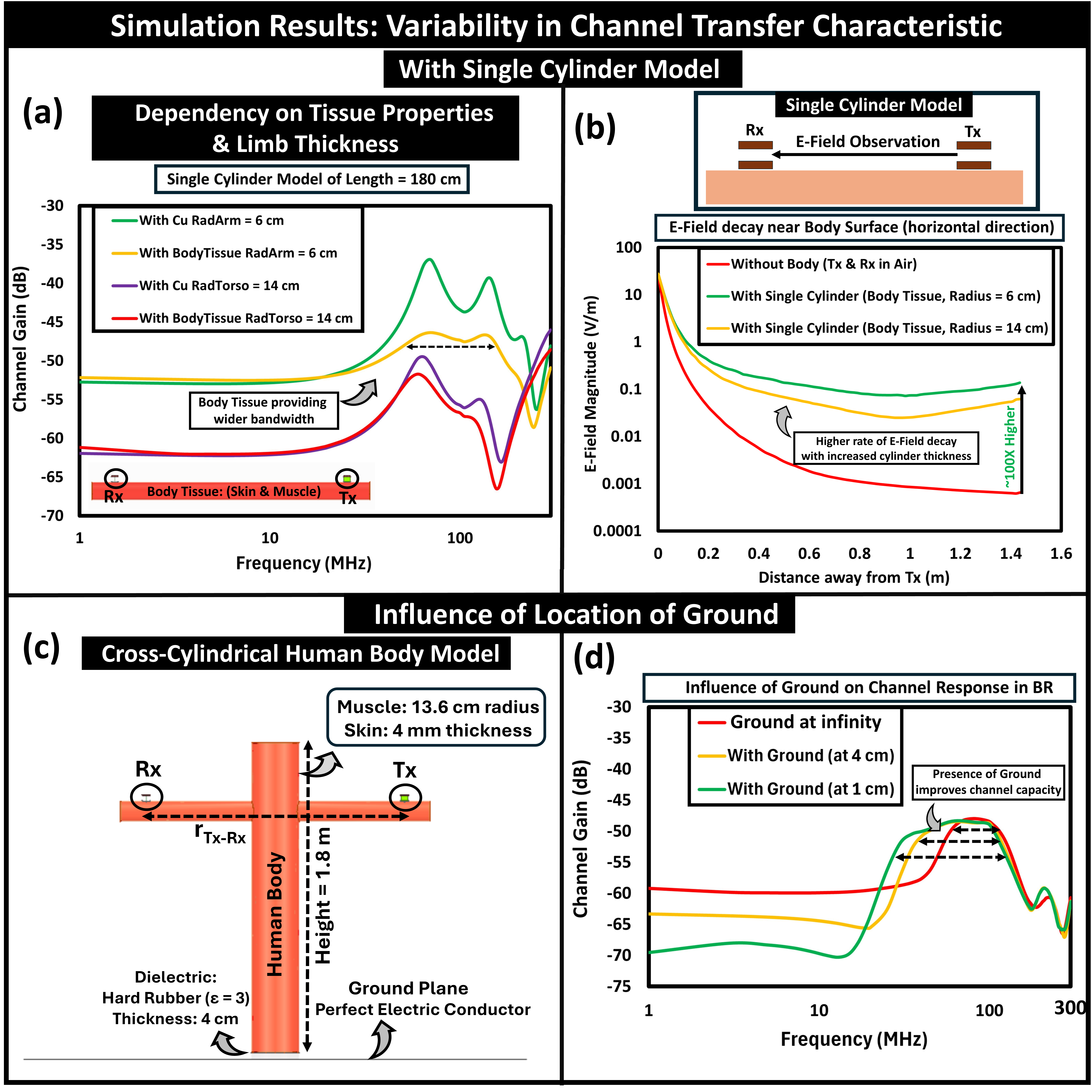}
\caption {\textbf{Analysis of Channel Transfer Characteristics:} \textbf{With Single Cylinder Model:} \textbf{a.} Influence of dielectric properties of body tissues and limb thickness on channel gain variability. \textbf{b.} E-Field decay profile over distance from the transmitter along the body surface. \textbf{Influence of Ground Location:} \textbf{c.} Numerical simulation setup with cross-cylindrical human body model (Front view). \textbf{d.} Variation in ground location relative to the subject's body affects operational bandwidth and channel capacity.}
\label{fig5}
\end{figure}

\subsubsection*{Influence of Body Tissues $\&$ Limb Thickness}
To understand how the tissue properties affect the behavior of the body channel, we conducted an analysis using a single-cylinder model. The transfer characteristic shown in Fig. \ref{fig5} (a) illustrates the dependence of channel gain on the dielectric properties of tissues (skin and muscle) and compares it with a conducting cylinder made up of copper. While body tissues, as imperfect electrical conductors, provide relatively lower channel gain, they can resonate with a lower quality factor from their inherent dielectric losses, allowing for a broader bandwidth over a conducting cylinder. We also observed that an increase in the radius of the cylinder, mimicking the thickness of a torso, led to a reduction in channel gain due to the thicker cylinder resulting in a sharper rate of attenuation in the observed E-field strength along the surface of the cylinder, as shown in Fig. \ref{fig5} (b). Furthermore, there was about a 100-fold improvement in the E-field strength along the body surface compared to device-to-device coupling, validating the previously observed channel gain benefits and providing motivation for analyzing BR-HBC.

\subsubsection*{Location of Ground influencing Body Channel Capacity}
It is crucial to analyze the subject's body position on the ground to optimize the body's signal transmission capacity for capacitive HBC. The numerical simulation setup to capture the change in the ground position about the subject's body is presented in Fig \ref{fig5} (c). In the context of EQS, being close to the ground reduces the channel gain by influencing the load impedance from the body (Z$_B$ = $\frac{1}{sC_B}$) by amplifying the parasitic capacitive coupling (C$_B$) between the human body and the earth's ground, as V$_{out}$ $\propto$ $\frac{1}{C_B}$. Additionally, it shifts the zero of the channel transfer function to a lower frequency, and hence, reduces the EQS flat band range, i.e., the operational bandwidth. However, in the BR frequency regime, the human body acts as a lossy resonator, and being closer to the ground causes the transfer function second pole to shift to a higher frequency. This results in a enhancement in operational bandwidth and thereby improvement in channel capacity, as illustrated in Fig. \ref{fig5} (d).

\begin{figure}[ht!]
\centering
\includegraphics[width=\linewidth]{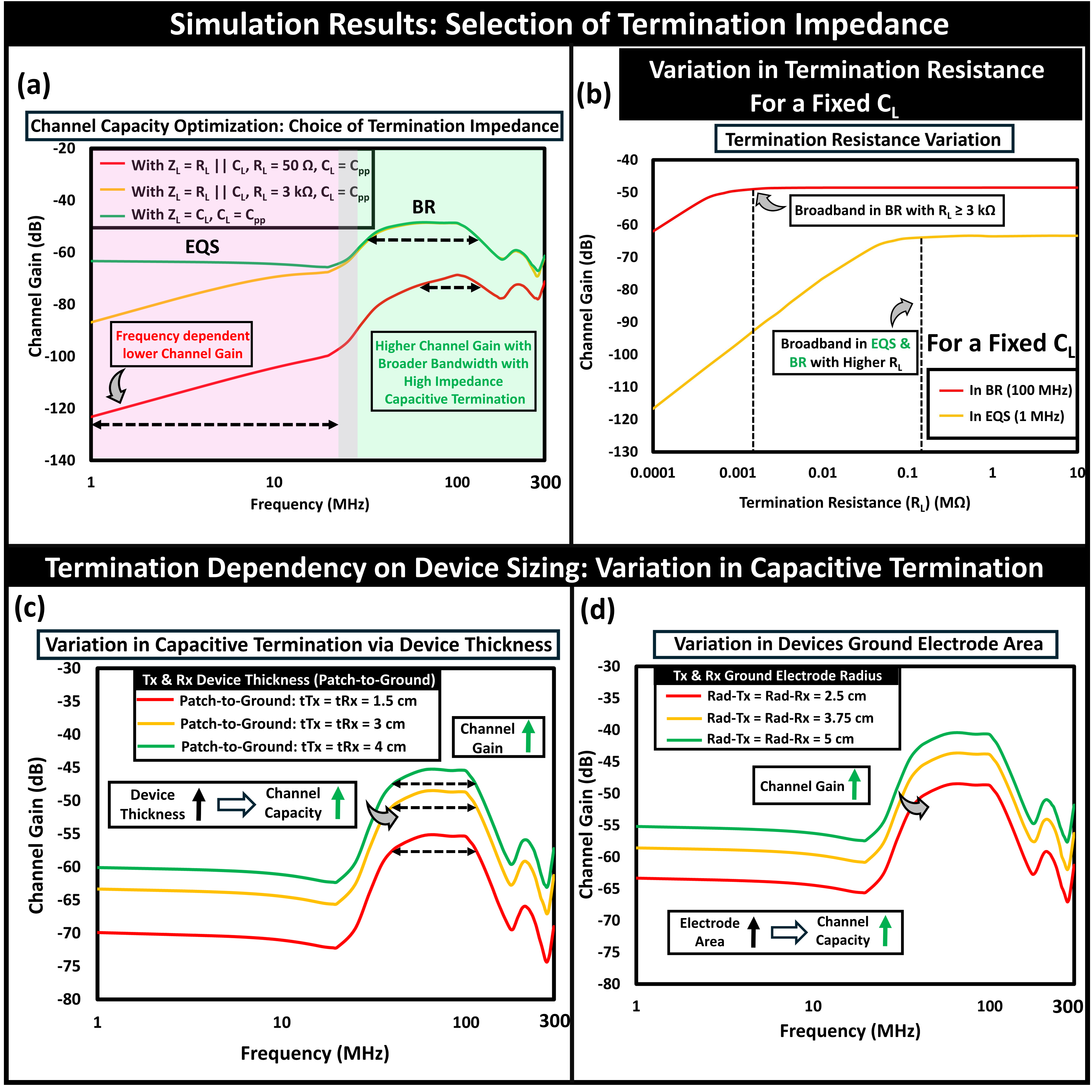}
\caption {\textbf{Choice of Termination Impedance:}
\textbf{a.} Variation in channel gain occurs with different types of termination, such as resistive, capacitive, or a combination of both. With a choice of high-impedance capacitive termination, broadband channels ranging from EQS to BR regime can be enabled. \textbf{b.} Variation in channel gain with a change in termination resistance. \textbf{Influence of Device Sizing:}
\textbf{c.} An increase in device thickness reduces the effective termination impedance at the receiver, thereby improving the channel gain. \textbf{d.} Increasing the ground electrode area of the devices increases the effective return path impedance at the receiver, resulting in an improved channel capacity.
}
\label{fig6}
\end{figure}

\subsubsection*{Selection of Termination Impedance}
The choice of termination impedance (Z$_L$) is critical for enhancing the efficiency of high-speed body-centric communication with wearable devices. It has a notable impact on the channel gain and operational bandwidth, as illustrated in Fig. \ref{fig6} (a). In voltage mode communication, a system with effective low-impedance resistive termination (Z$_L$ = R$_L$ $\parallel$ C$_L$,  where R$_L$ = 50 $\Omega$, C$_L$ = 2.3 pF) provides lower channel gain in the EQS regime that varies with frequency at 20 dB/decade. While this system offers relatively higher channel gain for a wider bandwidth in the BR regime compared to EQS, it is not the optimal choice for channel capacity maximization. This system can further improve its channel capacity in the BR regime by using a high impedance termination (i.e., Z$_L$ = R$_L$ $\parallel$ C$_L$, where R$_L$ $\geq$ 3 k$\Omega$, C$_L$ = 2.3 pF). For a fixed C$_L$ of 2.3 pF, the impact of varying the receiver's load resistance ($R_L$) at operating frequencies in EQS and BR is depicted in Fig. \ref{fig6} (b). However, using a higher resistive load at the receiver will increase the contribution of thermal noise voltage at higher frequencies and eventually degrade the Signal-to-Noise Ratio at the output. 
By using a high impedance purely capacitive termination (i.e., Z$_L$ = $\frac{1}{j\omega C_L}$) at the receiver, channel capacity can be enhanced over a \textbf{wide-band channel spanning from a few tens of kHz to hundreds of MHz}, while also making the channel gain in EQS regime frequency independent. The size of the communicating devices may also influence the channel capacity as they modify the effective termination reactance (X$_{c(eff)}$) from the parasitic coupling between their ground and the human body. Increasing the thickness (signal-to-ground separation) of the devices leads to improved channel capacity due to the rise in channel gain resulting from the increased effective load impedance (Z$_{L(eff)}$) at Rx, as shown in Fig \ref{fig6} (c). Moreover, a larger area of the ground electrode of the devices can provide improvement in the channel gain (4X increase in area of the Tx and Rx ground resulting in $\sim$2.5X increment in channel gain), presented in Fig \ref{fig6} (d).

\begin{figure}[ht!]
\centering
\includegraphics[width=\linewidth]{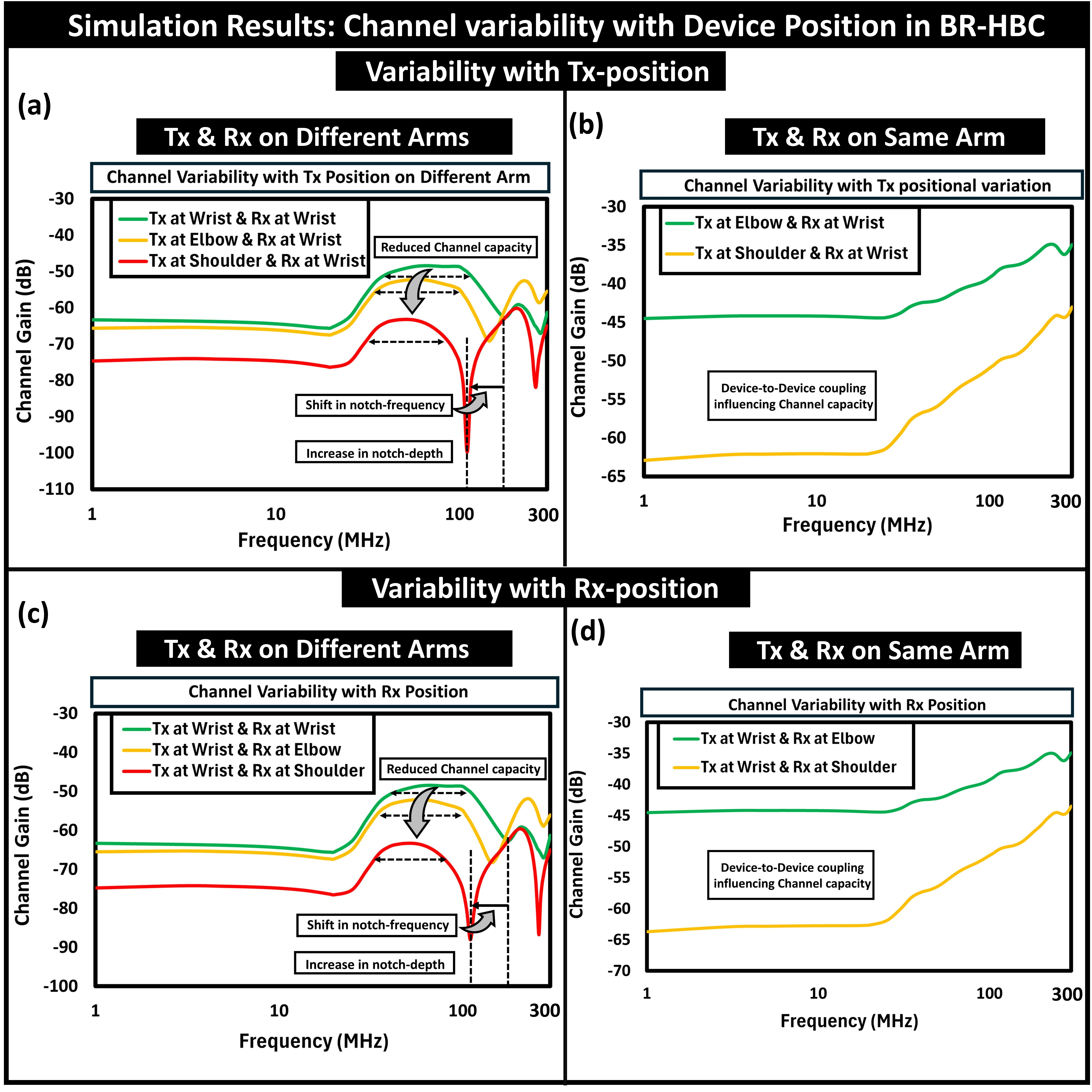}
\caption {\textbf{Device position influencing variability in channel capacity:} \textbf{Tx position variation}: \textbf{a.}Tx and Rx on different arms. \textbf{b.} Tx and Rx on the same arm. While on the opposite arms, the Tx movement towards the subject's torso reduces the channel capacity by attenuating the channel gain and operational bandwidth resulting from the combined effect of shadowing of field lines and notch movement in the channel characteristic to a lower frequency.  \textbf{Rx position variation}: \textbf{c.} Rx movement on the different arm. \textbf{d.} Rx movement on the same arm. Similar attributes in the channel variability that are observable with Tx movement along arms are noticeable with the positional change of Rx.}
\label{fig7}
\end{figure}

\subsubsection*{Factors contributing towards channel variability}
The on-body positioning of the communicating devices, i.e., the location of the signal feed (Tx position) and pickup (Rx position) from the surface of the human body, alters the formed EM-wave patterns and thereby leads to variability in channel characteristics, portrayed in Fig. \ref{fig7}. While keeping the Rx fixed at the wrist, the positional change of Tx along the opposite arm towards the shoulder reduces the channel capacity as it reduces the channel gain and operational bandwidth (i.e., the location of the first notch starts moving to a lower frequency), presented in Fig. \ref{fig7} (a). The channel gain attenuation happens as the field lines emanating from the Rx-ground get blocked from a conducting surface like the subject's body, i.e., an equivalent effect, conceptualized and termed body shadowing in EQS. Furthermore, the location of the notches from the first resonance moves to a lower frequency due to the higher relative permittivity of the body tissue (i.e., $\epsilon_{tissue}$ $>>$ $\epsilon_{air}$) 
The shift in the Tx position towards Rx on the same arm shows an improved channel gain owing to its behavior dominated by the device-to-device coupling, shown in Fig. \ref{fig7} (b). The movement of the Rx along the different and same arms also exhibit similar trends, captured in Fig. \ref{fig7} (c, d).

\begin{figure}[ht!]
\centering
\includegraphics[width=\linewidth]{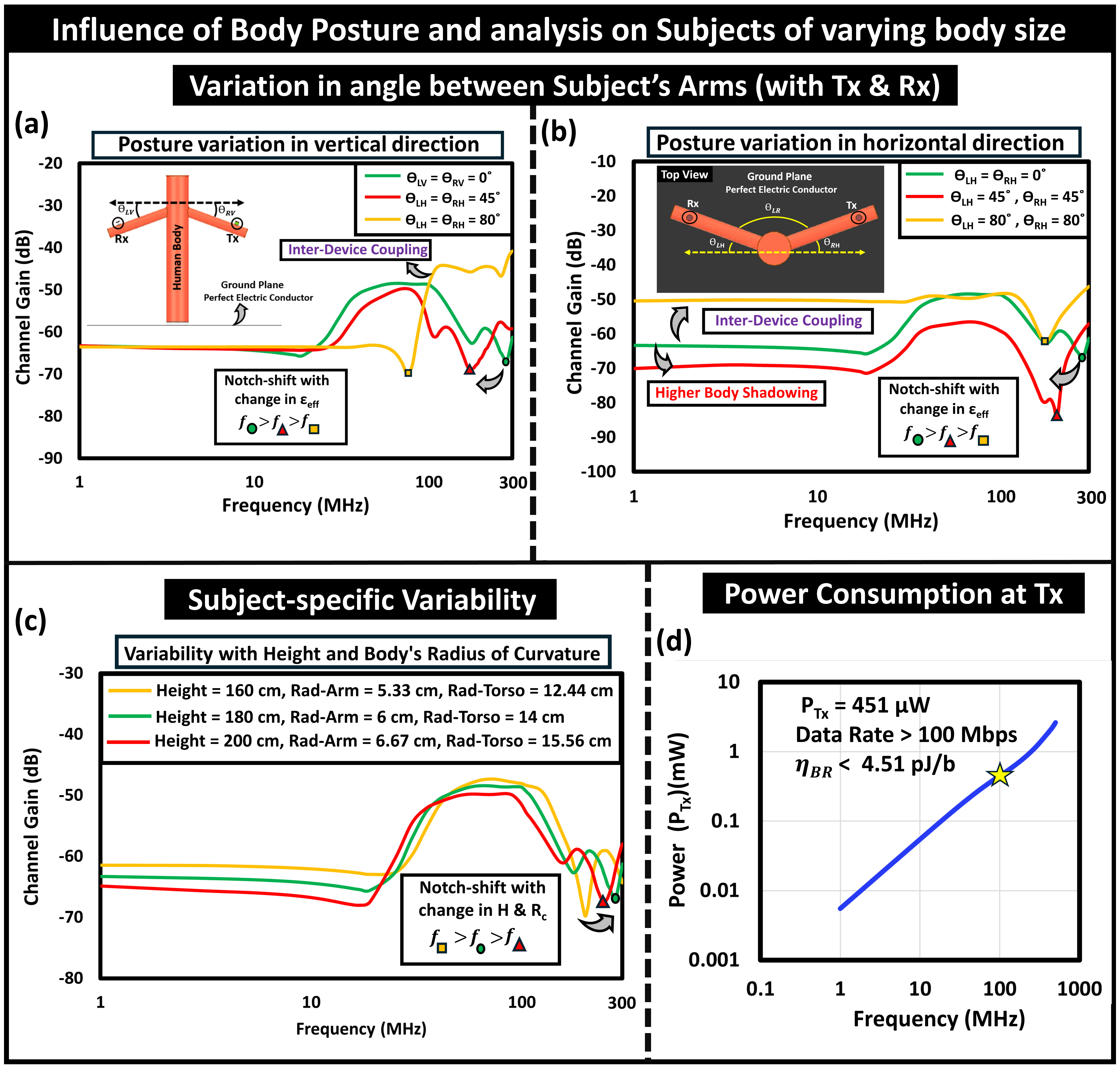}
\caption {\textbf{Subject's Body Posture influencing Channel Variability:} \textbf{Variation in angle between subject's arms}: \textbf{a.} Arms rotation in vertical direction. \textbf{b.} Arms rotation in horizontal direction. While keeping the devices in the opposite arms, the arm movement towards the subject's torso results in a notch shift in the channel characteristic to a lower frequency. \textbf{c.} Effect of change in subject's height and radius of curvature of arms and torso. \textbf{d.} Variation in consumed power at the transmitter highlighting the energy-efficiency of the high-speed communication link.}
\label{fig15}
\end{figure}

In addition to device positioning, changes in the subject's body posture also affect the variability of the channel characteristics. When the on-body locations of the transmitter (Tx) and receiver (Rx) are fixed, the movement of the arms to simulate different postures alters channel behavior, as illustrated in Fig. \ref{fig15}. For example, if the subject rotates arms vertically from a T-pose, the devices move closer to the torso, resulting in reduced channel gain and a shift in the notch frequency to a lower value, conversely, when the arms are held very close to the torso (i.e., $\theta_{LV}$ = $\theta_{RV}$ = 80$^\circ$), the inter-device coupling becomes more dominant, improving channel gain, as shown in Fig. \ref{fig15} (a). A similar trend in notch movement occurs with variations in horizontal posture, as presented in Fig. \ref{fig15} (b). The position of the notch depends on the effective permittivity ($\epsilon_{eff}$) encountered along the signal path, which also relies on the line of sight (LoS) between the communicating devices (Tx $\&$ Rx). Therefore, this relationship varies with changes in body posture. The effect of variation in the subject's body dimension on channel capacity is captured while keeping the aspect ratio of the human body model consistent, depicted in Fig. \ref{fig15} (c). With the increase in subject height in proportion to the radius of curvature of arms and torso, a decrease in channel gain in EQS and a shift in the notch to a lower frequency is observed. While analyzing the change in power consumption at the transmitter with the operating frequency increasing from EQS to BR regime and beyond, presented in Fig. \ref{fig15} (d), the viability of energy-efficient ($\eta <$ 4.5 pJ/bit) communication at 100's of Mbps via BR HBC is captured. Besides, the increasing trend of the consumed power confirms the increasing radiative nature of the energy flux from the transmitter at an operating frequency beyond BR frequency regime.

\begin{figure}[ht!]
\centering
\includegraphics[width=\linewidth]{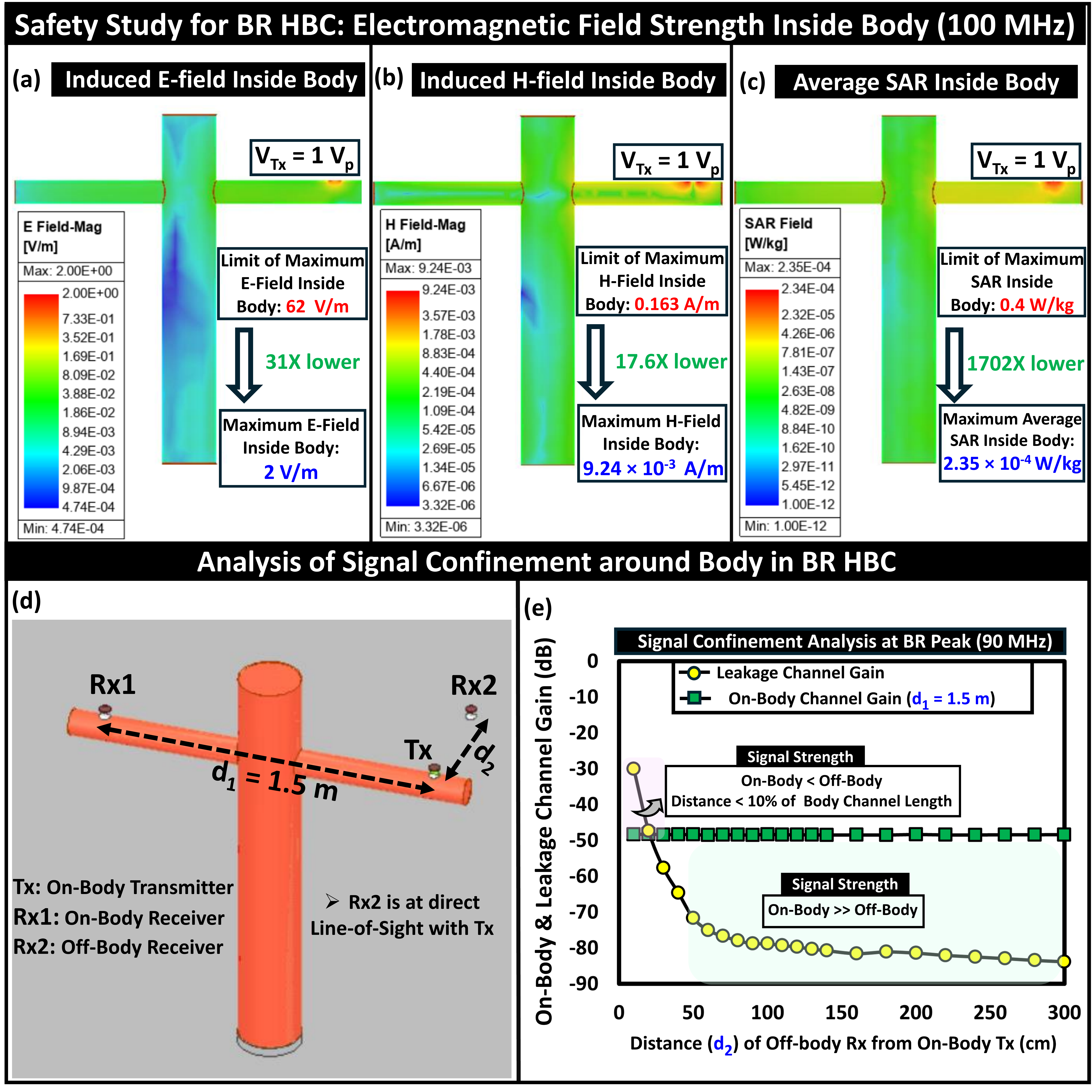}
\caption {\textbf{Analyzing safety aspects and leakage profile of BR HBC:} \textbf{a.} Induced E-Field remains an order of magnitude lower than the specified safety thresholds from ICNIRP. \textbf{b.} Induced H-Field stays several orders below the safety limits.
\textbf{c.} Average Specific Absorption Rate (SAR) lying below the upper threshold. \textbf{d.} Comparison of the on-body received signal to the off-body leakage signal.
}
\label{fig20}
\end{figure}

\subsubsection*{User's Safety}
The increase in the consumed transmitter current in the BR frequency regime over EQS demands an analysis of the safety aspects of deploying this technology ubiquitously. In a couple of previous studies \cite{maity2020safety, datta2024S2C}, the safety aspects of individuals using EQS-HBC and the recommended upper thresholds for human exposure to radio frequency (RF) electromagnetic fields are discussed as per the guidelines issued by the Institute of Electrical and Electronics Engineers (IEEE) \cite{ieee1992ieee} and the International Commission on Non-Ionizing Radiation Protection (ICNIRP) \cite{ICNIRP2020}. With its peak in channel transfer characteristic appearing between 50 to 150 MHz for subjects of height ranging from 160 to 190 cm, for the BR HBC, the upper thresholds are considered to be the induced field and specific absorption limits specified by ICNIRP. We take a numerical simulation-based approach to estimate the induced field strength (electric and magnetic) inside the human body and the Whole-Body Average Specific Absorption Rate (SAR), presented in Fig. \ref{fig20}. We performed electromagnetic analysis on the cross-cylindrical human model in HFSS. The estimated value of the \textbf{induced E-field} inside the body being \textbf{31X lower} than the maximum allowed E-field (in Fig. \ref{fig20} (a)), the \textbf{induced H-Field} being \textbf{17.64X lower} than the maximum H-field limit (in Fig. \ref{fig20} (b)), and the \textbf{whole-body average SAR} turning out to be \textbf{1702X lower} than the specified upper limit (in Fig. \ref{fig20} (c)) at an input voltage of amplitude 1 V \textbf{corroborates the operational safety of BR HBC}.

\subsubsection*{Signal Confinement Analysis}
In this section, we investigate the signal confinement behavior of the proposed communication link. The leakage profile around the subject's body for the proposed communication link during the line-of-sight scenario between on-body Tx and off-body Rx is shown in the Fig. \ref{fig20} (d, e). It is observed that at the BR peak (90 MHz), the signal strength received off the body (V$_{OffBody}$) remains comparable to or higher than the on-body signal strength (V$_{OnBody}$) when the distance between the on-body Tx and the off-body Rx becomes less than 15-20$\%$ of the fixed on-body channel length spanning 1.5 m. Its $\frac{V_{OffBody}}{V_{OnBody}}$ being $<$ 0.1 for an off-body distance of 0.5 m makes the signal confinement of BR HBC comparatively better than traditional radiative wireless communication techniques, where the leakage signal mostly remains comparable to or higher than the received signal strength on the body.

\begin{figure}[ht!]
\centering
\includegraphics[width=\linewidth]{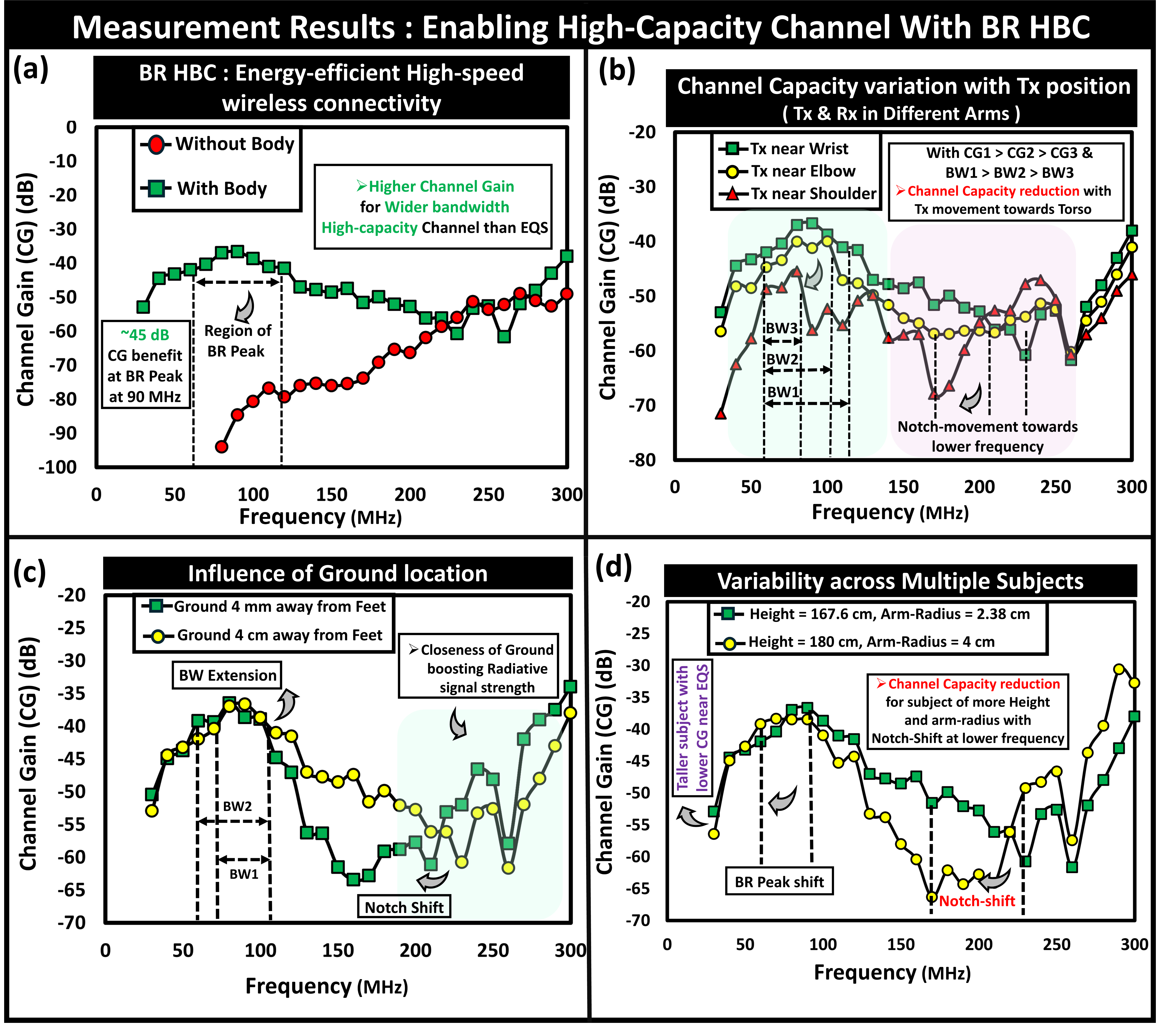}
\caption {\textbf{Experimental Results: High capacity body-centric communication channel and its variability in BR frequency regime:} \textbf{a.} BR Human Body Communication offering higher channel path gain for wider bandwidth to enable high-speed connectivity. \textbf{b.} Reduction in body channel capacity with change in Tx position along subject's arm towards torso. \textbf{c.} Influence of location of Ground. \textbf{d.} Influence of subject's body dimension on channel behavior through study on multiple subjects.}
\label{fig10a}
\end{figure}

\subsection*{Experimental Results}
To verify the trends observed in numerical simulations, we performed measurements on human subjects using wearable devices to demonstrate BR HBC. We use a coaxial SMA connector to couple the signal from the wearable transmitter to the wristband coupler. The coupler's surface, remaining in contact with the skin, is coated with double-sided conductive copper foil tape in a rectangular shape to emulate the signal patch. Subsequently, we picked up the signal at the receiving end by utilizing another coaxial SMA connector and a similarly designed wristband coupler to the wearable receiver. It's important to note that the signal couplers are not impedance matched to the human body's input impedance and, hence reflections observed in experiments may differ from the numerical simulation results at higher frequencies. Nevertheless, we can demonstrate that the human body is a high-capacity communication channel for high-speed body-centric communications. With the setup above, we conducted various experiments under different scenarios. 

The experimental verification, illustrated in Fig. \ref{fig10a} (a) shows the feasibility of using the human body as a energy-efficient, high-capacity wireless channel. It supports the theoretical understanding and confirms the channel behavior observed in Fig. \ref{fig4} (a) through numerical simulations. 
The changes in transmission location on the body while keeping the receiver fixed at the subject's wrist are shown in the experimental results in Fig. \ref{fig10a} (b). While moving the transmitter towards the subject's torso in the opposite arm to that of the arm having the receiver, a decrease in body channel capacity in the BR frequency regime is observed. These trends align with the simulation results, depicted previously in Fig. \ref{fig7} (a).
The dependency of body channel capacity on the ground location is portrayed in Fig. \ref{fig10a} (c), which is consistent with the results depicted in Fig. \ref{fig5} 
(d) as the closeness of ground boosts the high-frequency radiation.  

\subsubsection*{Channel Variability Across Subjects}
The measurements of channel gain in the BR frequency regime were conducted on various subjects over a long period. The results shown in Fig. \ref{fig10a} (d) illustrate the variation in channel characteristics across subjects with heights ranging from 165 cm to 180 cm and weights between 60 kg to 80 kg. With the increase in the height and arm-radius of the subject, the channel gain goes down, and the location of the body-resonance peak shifts towards lower frequencies. Furthermore, a noticeable shift in the notch-frequency is observed with increase in arm-radius which remains concurrent with the simulation results depicted in Fig. \ref{fig15} (c). Moreover, a taller subject with more arm-radius while acting as a bigger ground to the devices boosts the high-frequency radiation characteristics.  
The trend in the high-frequency undulations in the channel characteristic is confirmed through repeated measurements. 

\begin{figure}[ht!]
\centering
\includegraphics[width=\linewidth]{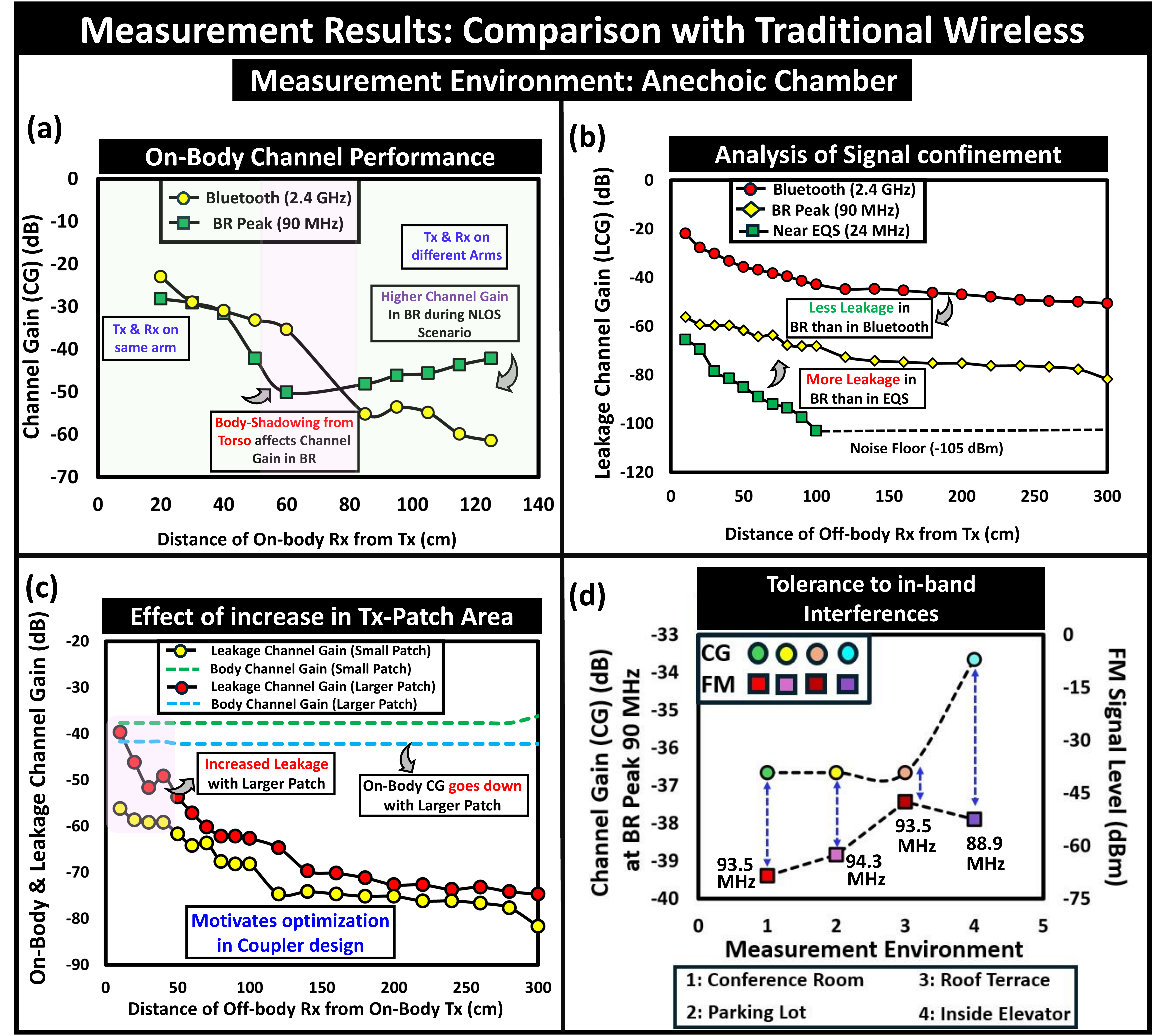}
\caption {\textbf{Experimental Results: Comparative study of proposed BR HBC with traditional antenna-based wireless communication inside Anechoic Chamber:} \textbf{a.} On-body channel performance with variation in on-body Rx position: Better channel gain with BR HBC during Non-Line-Of-Sight scenarios. \textbf{b.} Analysis of signal confinement between Bluetooth, BR and near EQS (24 MHz). \textbf{c.} Effect of increase in Tx-patch area on the on-body signal and leakage performance. \textbf{d.} Tolerance to in-band interference in different environments.}
\label{fig11}
\end{figure}

\subsubsection*{Comparison with Traditional Wireless}
We compared our proposed BR-HBC mode with traditional wireless communication like Bluetooth through two experiments: one studying on-body channel performance and the other investigating off-body leakage characteristics inside anechoic chamber environment. For the conventional wireless communication experiments, we used Bluetooth antennas connected to a handheld RF signal generator as the radio transmitter and a TinySA Ultra spectrum analyzer as the radio receiver in an anechoic chamber environment. During the on-body channel measurements, the test subject maintained a T-pose, holding the transmitter in one arm. In contrast, the on-body receiver's position varied from the wrist of the other arm to the near wrist of the arm where the transmitter was positioned, presented in Fig. \ref{fig10b} (a). The on-body channel gain showed distance-dependent attenuation characteristics away from the transmitter and sharper attenuation during non-line-of-sight scenarios between the transmitter and the receiver, illustrated in Fig. \ref{fig11} (a). 
With the BR HBC, although the on-body channel gain experiences higher attenuation from the body shadowing near the subject's torso, it shows better gain over antenna-based radiative communication during non-line-of-sight scenarios.
Off-body leakage analysis was conducted in a line-of-sight scenario between the on-body transmitter and the off-body receiver (placed on a wooden stool setup), shown in Fig. \ref{fig10b} (b). With the radiative nature of Bluetooth, the leakage channel gain remained at a level of approximately -50 dB over 3 meters from the transmitter, as shown in Fig. \ref{fig11} (b). The off-body leakage signal measurement for BR HBC was performed by keeping the RF-spectrum analyzer on a similar stool setup. The proposed BR-HBC showed faster attenuation of the off-body leakage characteristics away from the transmitter, and its signal level was approximately 20 dB lower than the on-body signal strength, making it a energy-efficient modality over radiative transmission, illustrated in Fig. \ref{fig11} (b).

Furthermore, while studying the influence of the coupler area on the relative strength of on and off-body signals, using a larger patch that follows the contour of the subject's arm, a sharper rise in the off-body channel gain was observed at the expense of slight reduction on-body signal strength (resulting from the increase in effective overlapping area between the signal and ground of the device), depicted in Fig. \ref{fig11} (c). This opens up the scope for optimizing the transmitter design, leading to the emergence of optimized form factor devices.

\subsubsection*{Tolerance to Interference}
With its electrically conducting nature, in the BR frequency regime, when the human body starts to resonate, it picks up the electromagnetic signals from the interference sources that are present in the surroundings. When the power level of these interferences exceeds the power of the operating signal (i.e., a lower value of Signal-to-Interference Ratio (SIR)), it may cause unreliable reception of digital data packets with a higher bit error rate. With its peak frequency residing in the range of 50 MHz to 150 MHz, the data transmission at the peak of the Body-Resonance can be susceptible to environmental interference in this frequency range, especially from the presence of an FM radio band that ranges from 87.9 to 107.9 MHz in the United States and is divided into 101 channels of 0.2 MHz wide. The variation in the level of measured FM signal level on the subject's body in comparison to the channel gain at the peak of BR-HBC under different scenarios (indoor and outdoor) are captured in Fig. \ref{fig11} (d). The on-body channel gain remaining significantly higher than the FM signal level (higher SIR), affirms the  interference tolerant nature of the proposed link. It is to be noted that an increase in the FM signal level in some scenarios (especially in case 3: Roof Terrace or close to any FM transmitter) if observed, which if degrades SIR, the techniques towards mitigating interferences can be incorporated. Hence, to fully utilize the higher channel capacity of the BR HBC, previously proposed techniques for interference tolerance like Adaptive Frequency Hopping (AFH) \cite{cho200960}, direct sequence spread spectrum (DSSS), Code-Division Multiple Access (CDMA), and Integrating DDR Receiver as Notch Filter \cite{sen2016socialhbc} for strong interference suppression can be used in conjunction to handle in-band interference like FM. 

\section{Discussion}
Body-Resonance HBC offers energy-efficient faster connectivity among miniaturized battery-powered body-connected devices. Building up our understanding of complex electromagnetic propagation around the human body, we started our analysis with a single-cylinder model emulating human torso and limb. Subsequently, we provide numerical simulation on a complete human model and experimental results on human subjects, covering the maximum channel length ($\sim$ 2 m) for on-body communication. Instead of optimizing the power gain, which requires relevant frequency-dependent matching at the transmitter and the receiver ends, we focused on demonstrating the viability of utilizing the \textbf{human body as a high capacity channel to support $\sim$100s of Mbps body-centric communication}. With the high impedance capacitive termination at the receiver, we showed the feasibility of wideband communication, spanning from a few tens of kHz to hundreds of MHz. Considering practical use cases from the user's convenience perspective, we performed measurements with ESD wristbands, having small rectangular metal contacts (2.5 cm $\times$ 3 cm) coated with copper tape as signal couplers. Experimentally verifying the influence of the increased contact area of the signal patch of Tx, presented in Fig. \ref{fig10b} (c), we concluded that by keeping the polarisation pattern at the electrode similar, a larger patch doesn't give appreciable improvement in on-body channel capacity but significantly increases the off-body leakage at reduced distance between the on-body Tx and off-body Rx. The \textbf{design flexibility of the Tx patch, not being necessarily required to follow the full contour over the body} (that happens in most practical scenarios) enhances its scope of applicability and motivates the design of devices with optimized form factor. Application-specific coupler design for devices residing at different parts of the body for the energy-efficient coupling of EM waves to the body surface and adding tuned impedance matching networks to minimize undesired radiative energy loss from the devices as a future goal can potentially inspire the design of the energy-efficient transceivers.

\section{Conclusion}\label{sec13}
In this study, we propose BR-HBC, a communication method that utilizes the resonance phenomena of the human body to create a low-loss, wide-band, high-speed communication link for miniaturized battery-powered body area network (BAN) devices. The Body-Resonance frequency range, where the operating signal wavelength approaches the length of the body channel, leads to the formation of electromagnetic resonant patterns on the human body while enabling high-speed connectivity. The BR HBC exhibits approximately 15-20 dB lower transmission path loss across a wide operational bandwidth (hundreds of MHz), resulting in a \textbf{$\sim$30X improvement in channel capacity compared to EQS}. This method provides better signal confinement around the body than traditional antenna-based wireless radio communication methods, making BR-HBC a promising choice for energy-efficient high-speed wireless connectivity. We demonstrated the feasibility of enabling a low loss, wide-band channel for \textbf{high-speed ($\sim$100s of Mbps) body-centric communication}. BR HBC has the potential to open up numerous possibilities for diverse body-centric applications that enrich augmented living.

 \section{Methods}\label{sec11}
This section provides a detailed explanation of the methods used for our numerical simulation and experiments. This will help independent researchers to reproduce our results in the future.

\subsection*{Setup for FEM-based Numerical Simulations}
\subsubsection*{Numerical Simulator $\&$ Human Body Model}
\begin{figure}[ht!]
\centering
\includegraphics[width=\linewidth]{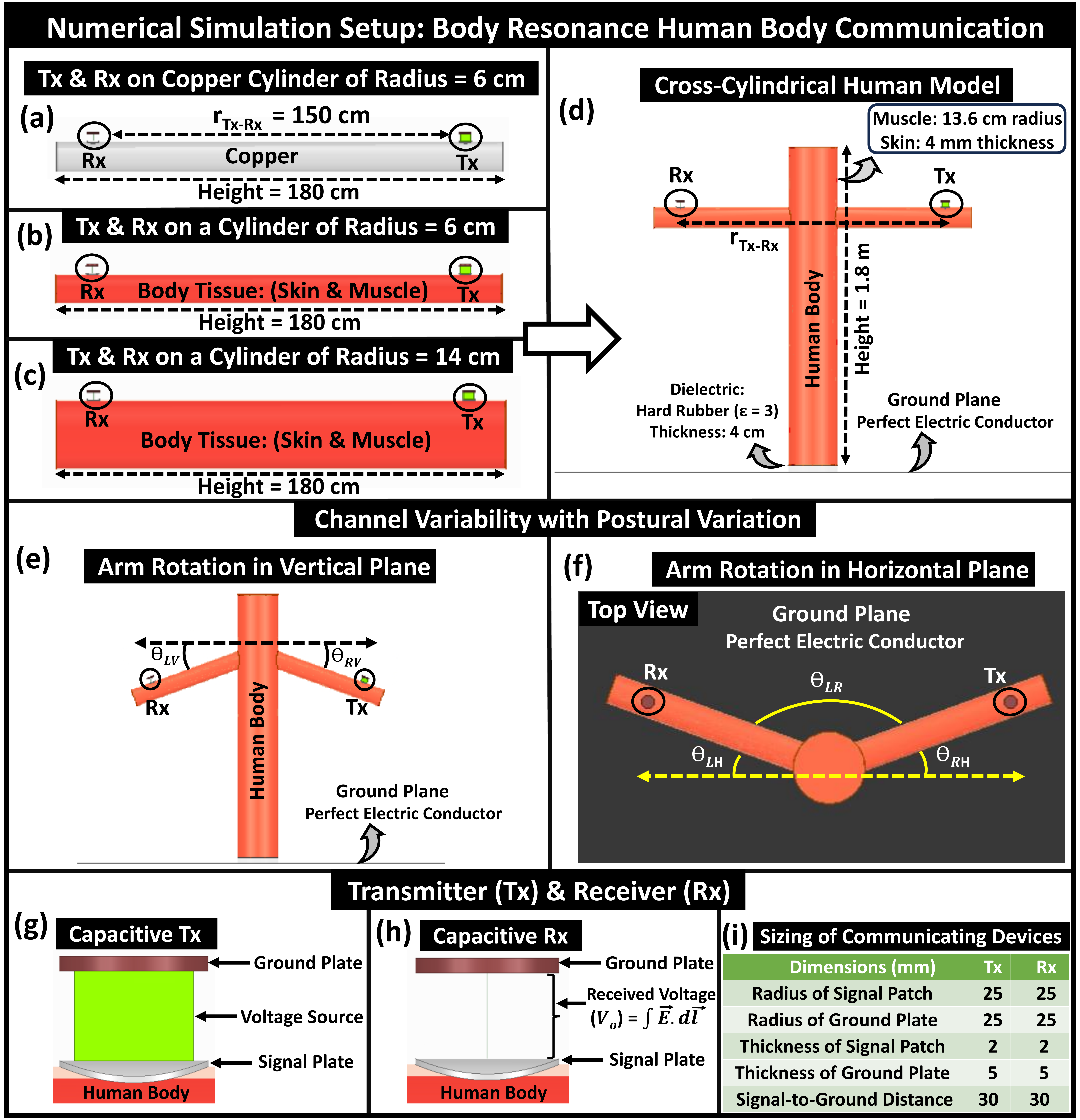}
\caption {\textbf{Setup for Finite Element Method (FEM)-based numerical simulation to enable Body-Resonance Human Body Communication (BR HBC):} \textbf{Simplified Single Cylinder Model:} with Tx and Rx on \textbf{a.}  Copper cylinder that replicates human arm of radius 6 cm. \textbf{b.} Body Tissue (skin and muscle) cylinder. \textbf{c.} Tissue cylinder with an increased thickness that emulates human torso of radius 14 cm. \textbf{d.} Cross-cylindrical human body model. \textbf{Subject's posture influencing channel characteristic}: Subject rotates its arms: \textbf{e.} in vertical plane (front view). \textbf{h.} in horizontal plane (top view). \textbf{Communicating Devices:} \textbf{g.} Wearable Transmitter (Tx) and \textbf{h.} Wearable Receiver (Rx) with their signal electrode/patch in contact with skin while respective ground electrode remains floating. \textbf{i.} Design specifications of the communicating devices.} 
\label{fig8}
\end{figure}

Electromagnetic simulations are performed using ANSYS High-Frequency Structure Simulator (HFSS), which uses Finite Element Method (FEM) to solve Maxwell's Equations with assigned boundary conditions. A simplified single-cylinder model (shown in Fig. \ref{fig8} (a, b, c)) and a cross-cylindrical model of the full human body using two perpendicular cylinders, each of height 180 cm with a radius of 14 cm and 6 cm representing respectively the torso and extended arms, is developed, illustrated in Fig. \ref{fig8} (d). The use of a simplified model has considerably reduced the computational time and complexity for numerical simulations, whose accuracy was validated previously through a comparison with a detailed model by Maity et al. \cite{maity2020safety}. The dielectric properties of (i.e., relative permittivity ($\epsilon_r$) and conductivity ($\sigma$)) of the body tissues (skin and muscle) have been taken from the works of Gabriel et al. \cite{Gabriel_1996}. The outer shells of the torso and arms are considered to be made up of skin of thickness 4 mm, with the interior muscle emulated by cylinders of radius 13.6 cm and 5.6 cm for the torso and arms. The arms of the cross-cylindrical model have been rotated at different angles to mimic the postural variation of the subject, as presented in Fig. \ref{fig8} (e, f). Replicating the scenario of a human standing on the ground, we assumed the human model is 4 cm above a plane with a Perfect E-Boundary assigned to it in HFSS that mimics an infinite ground plane similar to the earth's ground. 
With a solution type of modal network analysis and hybrid solver mode, the model is enclosed in an air region assigned to the Finite Element Boundary Integral (FEBI) boundary condition of the following dimensions: 168 cm $\times$ 292 cm $\times$ 296 cm. Instead of the commonly used radiation boundary which requires a bigger region, the FEBI boundary allows greater flexibility in choosing smaller-sized regions without sacrificing accuracy, leading to faster simulations.

\subsubsection*{Excitation}
A parallel-plate model for the capacitive transmitter, presented in Fig. \ref{fig8} (g), with the signal electrode connected to the body and the ground electrode, kept floating, is used to excite the human model. Two copper discs with radii of 2.5 cm are used to function as couplers. One of the discs of thickness 2 mm is curved onto the subject's body to mimic the signal patch, and another disc of thickness 5 mm is used to emulate the ground electrode of a wearable HBC device. The signal and the ground plates are separated at 3 cm. Emulating an AC voltage source of alternating potential difference of amplitude (V$_{in}$) 1 V, in HFSS, a voltage source excitation is placed between the signal and ground plates of the transmitter. Unlike the lumped port, suitable for 50 $\Omega$ matched excitation, the assigned voltage source, with its signal and ground plates, constitutes the transmitter (Tx) setup.

\subsubsection*{Calculation of Received Voltage}
The receiver, illustrated in Fig. \ref{fig8} (h), is designed to be structurally similar to that of the transmitter. The potential difference between the signal and the ground plate is calculated by integrating the electric field along a line, joining the two plates of the receiver.

\subsubsection*{Calculation of Transmitter Current $\&$ Power}
To measure the current consumption at the transmitter a lumped resistor (R) of 1$\Omega$ is connected in series with the voltage source excitation. Introducing a thin copper cylinder of thickness 0.5 mm and radius 2.5 cm at the midway between Tx-signal and ground, the peak voltage (V$_R$) across the 1$\Omega$ resistor (i.e., between the middle plate and signal patch) is measured to calculate the root mean square (r.m.s) value of the consumed current (I$_{Tx}$ = $\frac{V_R}{\sqrt{2}}$) at the transmitter. The power consumption at the Tx is calculated as $P_{Tx} = V_{Tx}I_{Tx}$ where $V_{Tx} = \frac{V_{in}}{\sqrt{2}}$ represents r.m.s value of applied input voltage (V$_{in}$). 


\begin{figure}[ht!]
\centering
\includegraphics[width=\linewidth]{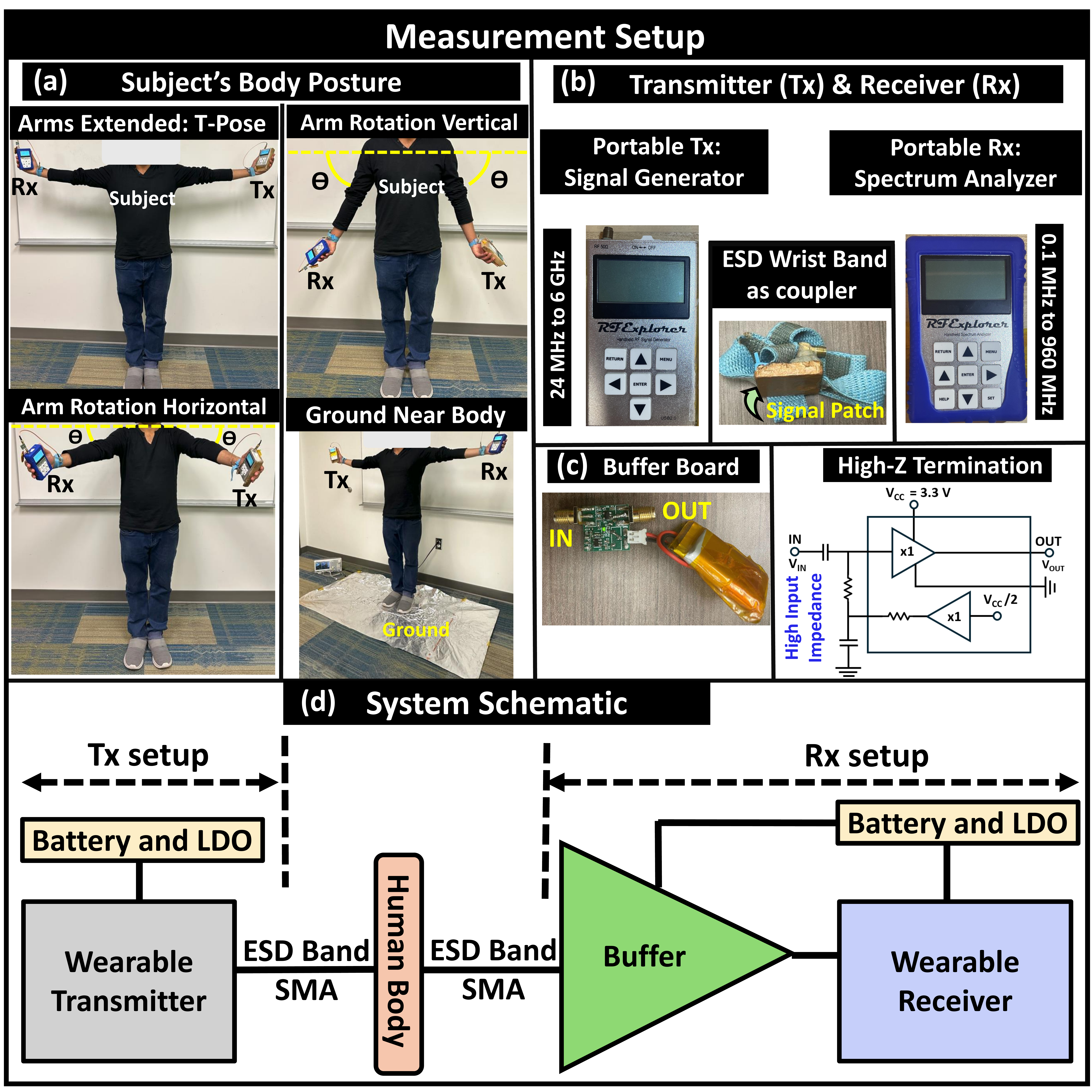}
\caption {\textbf{Measurement Setup:}  \textbf{Subject’s body posture for channel gain measurements:} \textbf{a.} The subject holds the portable transmitter (Tx) with its signal coupler in contact with the wrist of one arm, and the portable receiver (Rx) at the wrist of another arm, while emulating a T-pose and various postural variations to capture variability in channel capacity. \textbf{b.} Portable Signal Generator used as the transmitter. ESD wrist bands used as couplers to couple the signal to and pickup from the subject's body. Portable Spectrum Analyzer used as the receiver. The receiver setup includes the Tiny SA spectrum analyzer together with Buffer. \textbf{c.} Customized Buffer with high impedance capacitive termination for voltage mode communication. \textbf{d.} Schematic of the measurement system.}
\label{fig9}
\end{figure}

\begin{figure}[ht!]
\centering
\includegraphics[width=\linewidth]{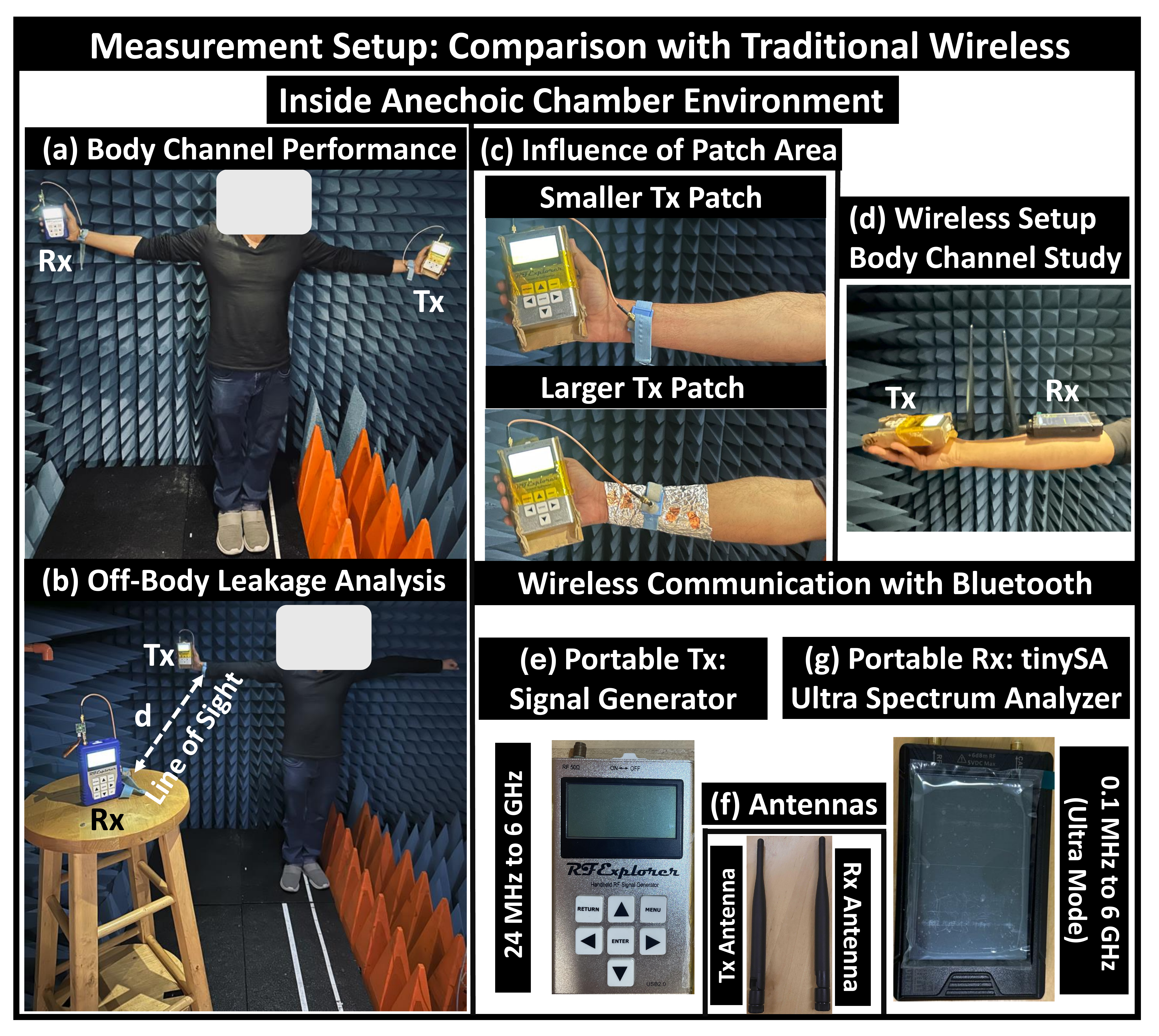}
\caption {\textbf{Measurement Setup:}
\textbf{Comparison of the BR HBC with traditional wireless inside Anechoic Chamber Environment:} \textbf{a.} Subject's body posture during body-channel performance study. \textbf{b} Setup for Off-Body leakage analysis with on-body Tx and Off-body Rx (placed at the Line-of-Sight of Tx). \textbf{c.} Influence of Tx patch area. \textbf{d.} Setup for wireless communication with Bluetooth. \textbf{e.} Portable signal generator as Tx. \textbf{f.} Tx and Rx antennas. \textbf{g.} TinySA Ultra as portable receiver.}
\label{fig10b}
\end{figure}

\subsection*{Experimental Setup}
This section portrays the setup used for channel gain and 
 leakage profile measurements. The protocols for experiments involving human subjects have been approved by the Purdue Institutional Review Board (IRB Protocol 1610018370). After obtaining informed consent from all participants, we performed experiments while complying with all the guidelines and regulations given by the Purdue IRB. The experiments are performed in two different environments: the first set of experiments, which analyzes the influencing factors towards body channel variability, are executed in a standard conference room, illustrated in Fig. \ref{fig9} and the second set of measurements is taken in a controlled environment to achieve better noise immunity such as an EM anechoic chamber, presented in Fig. \ref{fig10b}. Emulating the scenario of wearable-size communicating devices, battery-operated, handheld transmitter and receiver, shown in Fig. \ref{fig9} (b) are used for channel measurements. Instead of using a ground-connected Vector Network Analyzer (VNA) that offers an optimistic estimation of channel gain by keeping the grounds of the transmitter and receiver at the same potential, the employed handheld communicating devices closely replicate the realistic channel gain measurements.

\subsubsection*{Wearable Signal Transmitter}
A handheld RF signal generator (RFE6GEN) from RF, is employed as the signal transmitter that couples the signal to the body. With its operating frequency ranging from 24 MHz to 6 GHz and a resolution bandwidth of 1 kHz, this handheld signal generator (dimension: 11.3 cm $\times$ 7 cm $\times$ 2.5 cm) with its 50 $\Omega$ output impedance, is set at -5 dBm ($\sim$0.5 V peak-to-peak), which ensures that the in-body field and current densities to satisfy the ICNIRP safety limits \cite{ICNIRP2020}. A thin layer of cut-board casing is taped around the device to avoid any direct contact with the aluminum-made outer casing of the device that is connected to the device's ground floating ground of the transmitter.

\subsubsection*{Wearable Signal Receiver $\&$ Buffer}
A handheld spectrum analyzer (WSUB1G+) from RF Explorer, having an operating frequency ranging from 100 kHz to 960 MHz and a frequency resolution of 1 kHz, is used as the receiver. Similar isolation measures that were taken previously for the wearable transmitter are adopted for the receiver to prevent the subject from shorting to the receiver ground. However, this spectrum analyzer, with its 50 $\Omega$ input impedance, requires a buffer at its input for high-impedance capacitive terminated measurements. A buffer board is customized on a PCB using BUF602 (a high-speed, wide bandwidth buffer IC from Texas Instruments), presented in Fig. \ref{fig9} (c). A small-sized 3.7 V (nominal) rechargeable lipo battery powers up the buffer circuit. With its high bandwidth of 1 GHz and high slew rate of 8000 V/$\mu$s, it is suitable to handle high-speed AC signals. The high-impedance capacitive termination is ensured by the input impedance of the buffer, which is about 2 pF capacitive termination at the input side.

\subsubsection*{Couplers at the transmitting and receiving ends}
The functionality of the signal couplers to the body at the transmitting and receiving ends is emulated using ESD wristbands, which are connected to the devices via SMA cables and connectors. The length of the SMA cable is kept short to reduce excessive radiation from it, that can give rise to movement of the device peaks to a lower frequency leading to optimistic channel gain measurements.

\subsubsection*{Devices for measurements with Bluetooth}
In experiments with traditional wireless communication methods, such as Bluetooth, we utilize two 2.4 GHz omnidirectional antennas, shown in Fig. \ref{fig10b} (f). These antennas function within the frequency range of 2.4 - 2.485 GHz and possess linear vertical polarization, a gain of 6 dBi, VSWR less than 2.0, and an impedance of 50 ohms. We have connected one antenna to the RF signal generator to serve as the transmitter setup, while the other antenna is connected to the TinySA Ultra spectrum analyzer, which has a measuring range from 100 kHz to 5.3 GHz in its Ultra mode and is employed as a wearable receiver, illustrated in Fig. \ref{fig10b} (d, e, f, g).

\subsubsection*{Calibrations and Measurement Procedure}
\textbf{Calibration:}

To ensure accurate measurements with portable communicating devices, both the devices and the buffer need to be calibrated against wall-connected standard devices. The calibration of the devices and characterization of the buffer are performed using a benchtop signal generator and spectrum analyzer as standards, and the calibration process is summarized as follows:

\begin{itemize}
\item Handheld Transmitter: Connecting the handheld signal generator to a benchtop signal analyzer from Keysight for a fixed output power level of the transmitter the power level shown on the benchtop signal analyzer has been recorded, and its corresponding correction factor is calculated.
\item Handheld Receiver: Connecting the handheld receiver to a benchtop signal generator from Keysight, the frequency of the benchtop signal generator has been varied and the difference in power shown by the two devices is recorded. From the recorded power levels via interpolation, the receiver's correction factor at a specific frequency is obtained.
\item Buffer: The customized buffer is characterized by connecting its input to the signal generator and the output to the benchtop Keysight signal analyzer. The recorded difference between the shown power levels by the benchtop signal analyzer and the calibrated actual transmitted power is taken to calculate the correction factor from the buffer characterization.
\item Subsequently, the channel gain for each measurement is calculated from the recorded power shown by the handheld receiver by subtracting the transmitted power from it while taking the correction factors from the calibration of the devices and buffer into account.
\end{itemize}

\textbf{Measurement Procedure:}

The performance of the body channel has been experimentally studied in the BR regime by sweeping frequencies for specific positions of the transmitting and receiving devices on the body and by sweeping distances to capture the channel variability. Channel gain measurements in the absence of the human body were conducted by placing the transmitting and receiving devices at a certain distance on two wooden stools. The leakage profile around the subject's body was measured using a wearable transmitter worn on the body and keeping the handheld receiver on a wooden stool setup.

\backmatter

\section*{Supplementary Material}\label{sec12}
\subsection*{Supplementary Discussion 1: \\
Current Density and H-field Distribution:}
We present a comparative analysis of the current density and magnetic field distribution via numerical electromagnetic analysis between the EQS and BR frequency regimes, illustrated in Fig. \ref{fig13}. When the operating wavelength ($\lambda$) greatly exceeds the maximum dimension of on-body channels (i.e., $\lambda$ $>>$ l$_{Body}$ ) in the EQS regime, a consistent potential exists throughout the cylindrical body model, resulting in a uniform current distribution. Conversely, in the BR regime, where the wavelength $\lambda$ is comparable to the body channel length, there is an increased non-uniform current distribution within the conductor. This variability in current density inside the volume of the human body affirms its conceptual model as a lossy transmission, delineated in Fig. \ref{fig13} (b, e). Furthermore, the direction of the H-field indicates the direction of the current carried by the human body. In the EQS regime, the induced H-field is significantly lower than the induced E-field with an electric dipole, allowing us to disregard the influence of the H-field, shown in Fig. \ref{fig13} (c). Nonetheless, the shift in the direction of the H-field vector, as shown in Fig. \ref{fig13} (d, f), confirms the unbalanced, lossy transmission line nature of the human body, with the body as a signal conductor and the earth's ground acting as a ground conductor. We have included an animation plot illustrating the variation in H-field with frequency as Supplementary Movie 2 with this manuscript: \url{https://github.com/SparcLab/BodyResonanceHBC}

\begin{figure}[ht]
\centering
\includegraphics[width=\linewidth]{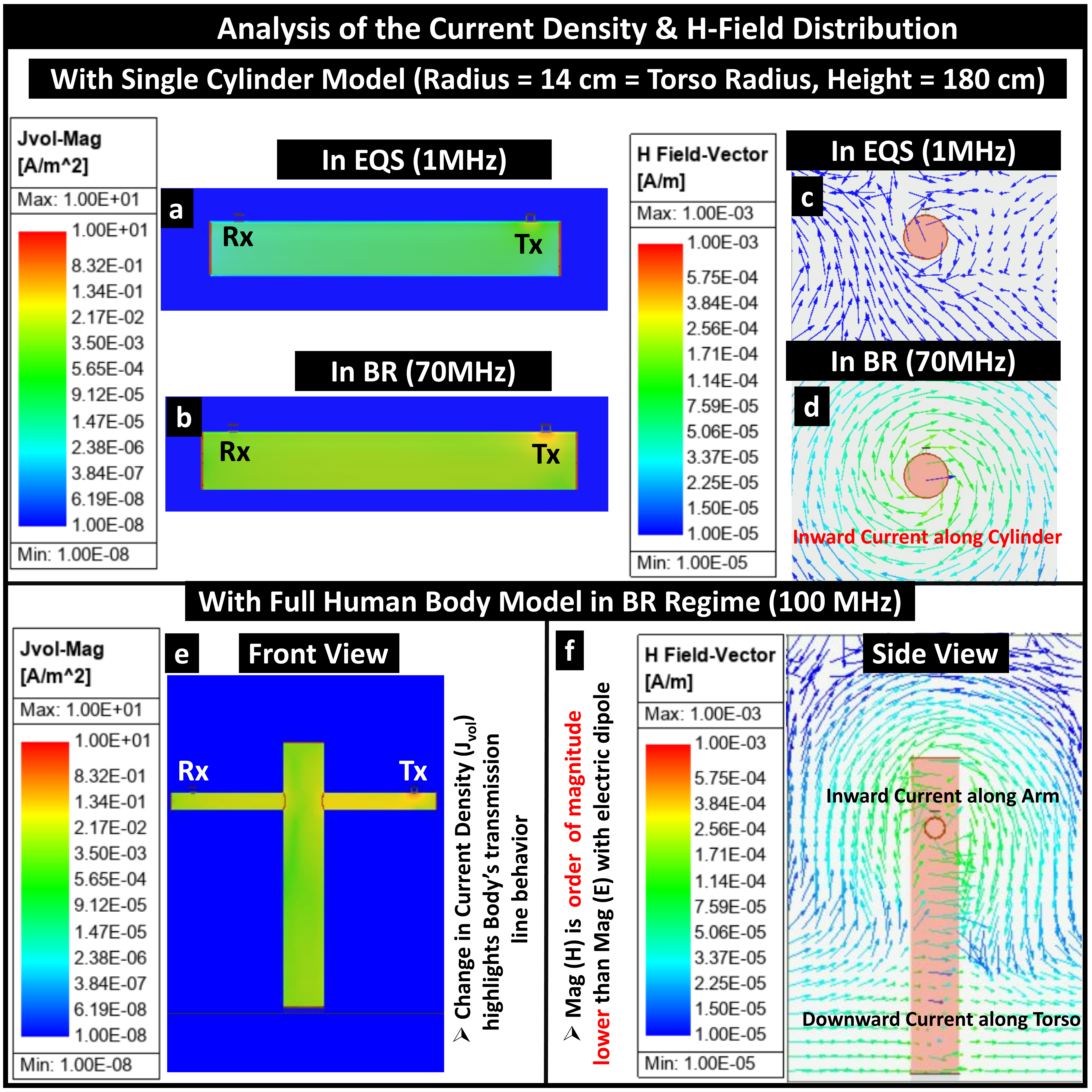}
\caption {\textbf{Supplementary Figure 1: Analysis of Current and H-Field Distribution: with Single Cylinder Model:}  The complex magnitude of the current density (Jvol) over the volume: \textbf{a.} In EQS, \textbf{b.} In Body-Resonance. Comparison of the magnetic field vector: \textbf{c.} In EQS. \textbf{d.} In BR. \textbf{with Cross-Cylindrical Human Body Model in BR at 100 MHz:} \textbf{e.} Current Density variation over the volume of the body. \textbf{f.} H-field vector.}
\label{fig13}
\end{figure}

\section*{Data Availability:}
The datasets that support the plots of numerical simulation and experimental results along with other findings that are presented in this paper, are available from the corresponding author upon reasonable request. Source datasets are presented in this paper.
\section*{Code Availability:}
Custom codes that are used to process the data are available from the corresponding author upon reasonable request.


\bigskip
\bibliography{sn-article}


\begin{thebibliography}{40}
\ifx \bisbn   \undefined \def \bisbn  #1{ISBN #1}\fi
\ifx \binits  \undefined \def \binits#1{#1}\fi
\ifx \bauthor  \undefined \def \bauthor#1{#1}\fi
\ifx \batitle  \undefined \def \batitle#1{#1}\fi
\ifx \bjtitle  \undefined \def \bjtitle#1{#1}\fi
\ifx \bvolume  \undefined \def \bvolume#1{\textbf{#1}}\fi
\ifx \byear  \undefined \def \byear#1{#1}\fi
\ifx \bissue  \undefined \def \bissue#1{#1}\fi
\ifx \bfpage  \undefined \def \bfpage#1{#1}\fi
\ifx \blpage  \undefined \def \blpage #1{#1}\fi
\ifx \burl  \undefined \def \burl#1{\textsf{#1}}\fi
\ifx \doiurl  \undefined \def \doiurl#1{\url{https://doi.org/#1}}\fi
\ifx \betal  \undefined \def \betal{\textit{et al.}}\fi
\ifx \binstitute  \undefined \def \binstitute#1{#1}\fi
\ifx \binstitutionaled  \undefined \def \binstitutionaled#1{#1}\fi
\ifx \bctitle  \undefined \def \bctitle#1{#1}\fi
\ifx \beditor  \undefined \def \beditor#1{#1}\fi
\ifx \bpublisher  \undefined \def \bpublisher#1{#1}\fi
\ifx \bbtitle  \undefined \def \bbtitle#1{#1}\fi
\ifx \bedition  \undefined \def \bedition#1{#1}\fi
\ifx \bseriesno  \undefined \def \bseriesno#1{#1}\fi
\ifx \blocation  \undefined \def \blocation#1{#1}\fi
\ifx \bsertitle  \undefined \def \bsertitle#1{#1}\fi
\ifx \bsnm \undefined \def \bsnm#1{#1}\fi
\ifx \bsuffix \undefined \def \bsuffix#1{#1}\fi
\ifx \bparticle \undefined \def \bparticle#1{#1}\fi
\ifx \barticle \undefined \def \barticle#1{#1}\fi
\bibcommenthead
\ifx \bconfdate \undefined \def \bconfdate #1{#1}\fi
\ifx \botherref \undefined \def \botherref #1{#1}\fi
\ifx \url \undefined \def \url#1{\textsf{#1}}\fi
\ifx \bchapter \undefined \def \bchapter#1{#1}\fi
\ifx \bbook \undefined \def \bbook#1{#1}\fi
\ifx \bcomment \undefined \def \bcomment#1{#1}\fi
\ifx \oauthor \undefined \def \oauthor#1{#1}\fi
\ifx \citeauthoryear \undefined \def \citeauthoryear#1{#1}\fi
\ifx \endbibitem  \undefined \def \endbibitem {}\fi
\ifx \bconflocation  \undefined \def \bconflocation#1{#1}\fi
\ifx \arxivurl  \undefined \def \arxivurl#1{\textsf{#1}}\fi
\csname PreBibitemsHook\endcsname

\bibitem{li2015internet}
\begin{barticle}
\bauthor{\bsnm{Li}, \binits{S.}},
\bauthor{\bsnm{Xu}, \binits{L.D.}},
\bauthor{\bsnm{Zhao}, \binits{S.}}:
\batitle{The internet of things: a survey}.
\bjtitle{Information systems frontiers}
\bvolume{17},
\bfpage{243}--\blpage{259}
(\byear{2015})
\end{barticle}
\endbibitem

\bibitem{kuzlu2015review}
\begin{bchapter}
\bauthor{\bsnm{Kuzlu}, \binits{M.}},
\bauthor{\bsnm{Pipattanasomporn}, \binits{M.}},
\bauthor{\bsnm{Rahman}, \binits{S.}}:
\bctitle{Review of communication technologies for smart homes/building applications}.
In: \bbtitle{2015 IEEE Innovative Smart Grid Technologies-Asia (ISGT ASIA)},
pp. \bfpage{1}--\blpage{6}
(\byear{2015}).
\bcomment{IEEE}
\end{bchapter}
\endbibitem

\bibitem{baker2017internet}
\begin{barticle}
\bauthor{\bsnm{Baker}, \binits{S.B.}},
\bauthor{\bsnm{Xiang}, \binits{W.}},
\bauthor{\bsnm{Atkinson}, \binits{I.}}:
\batitle{Internet of things for smart healthcare: Technologies, challenges, and opportunities}.
\bjtitle{Ieee Access}
\bvolume{5},
\bfpage{26521}--\blpage{26544}
(\byear{2017})
\end{barticle}
\endbibitem

\bibitem{habibzadeh2019survey}
\begin{barticle}
\bauthor{\bsnm{Habibzadeh}, \binits{H.}},
\bauthor{\bsnm{Dinesh}, \binits{K.}},
\bauthor{\bsnm{Shishvan}, \binits{O.R.}},
\bauthor{\bsnm{Boggio-Dandry}, \binits{A.}},
\bauthor{\bsnm{Sharma}, \binits{G.}},
\bauthor{\bsnm{Soyata}, \binits{T.}}:
\batitle{A survey of healthcare internet of things (hiot): A clinical perspective}.
\bjtitle{IEEE Internet of Things Journal}
\bvolume{7}(\bissue{1}),
\bfpage{53}--\blpage{71}
(\byear{2019})
\end{barticle}
\endbibitem

\bibitem{elhayatmy2018internet}
\begin{botherref}
\oauthor{\bsnm{Elhayatmy}, \binits{G.}},
\oauthor{\bsnm{Dey}, \binits{N.}},
\oauthor{\bsnm{Ashour}, \binits{A.S.}}:
Internet of things based wireless body area network in healthcare.
Internet of things and big data analytics toward next-generation intelligence,
3--20
(2018)
\end{botherref}
\endbibitem

\bibitem{sen2020body}
\begin{barticle}
\bauthor{\bsnm{Sen}, \binits{S.}},
\bauthor{\bsnm{Maity}, \binits{S.}},
\bauthor{\bsnm{Das}, \binits{D.}}:
\batitle{The body is the network: To safeguard sensitive data, turn flesh and tissue into a secure wireless channel}.
\bjtitle{IEEE Spectrum}
\bvolume{57}(\bissue{12}),
\bfpage{44}--\blpage{49}
(\byear{2020})
\end{barticle}
\endbibitem

\bibitem{chatterjee2023bioelectronic}
\begin{barticle}
\bauthor{\bsnm{Chatterjee}, \binits{B.}},
\bauthor{\bsnm{Mohseni}, \binits{P.}},
\bauthor{\bsnm{Sen}, \binits{S.}}:
\batitle{Bioelectronic sensor nodes for the internet of bodies}.
\bjtitle{Annual Review of Biomedical Engineering}
\bvolume{25},
\bfpage{101}--\blpage{129}
(\byear{2023})
\end{barticle}
\endbibitem

\bibitem{sen2024human}
\begin{bchapter}
\bauthor{\bsnm{Sen}, \binits{S.}},
\bauthor{\bsnm{Datta}, \binits{A.}}:
\bctitle{Human-inspired distributed wearable ai}.
In: \bbtitle{Proceedings of the 61st ACM/IEEE Design Automation Conference},
pp. \bfpage{1}--\blpage{4}
(\byear{2024})
\end{bchapter}
\endbibitem

\bibitem{garn1995present}
\begin{barticle}
\bauthor{\bsnm{Garn}, \binits{H.}},
\bauthor{\bsnm{Gabriel}, \binits{C.}}:
\batitle{Present knowledge about specific absorption rates inside a human body exposed to radiofrequency electromagnetic fields}.
\bjtitle{Health physics}
\bvolume{68}(\bissue{2}),
\bfpage{147}--\blpage{156}
(\byear{1995})
\end{barticle}
\endbibitem

\bibitem{rehman2010investigation}
\begin{barticle}
\bauthor{\bsnm{Rehman}, \binits{M.U.}},
\bauthor{\bsnm{Gao}, \binits{Y.}},
\bauthor{\bsnm{Wang}, \binits{Z.}},
\bauthor{\bsnm{Zhang}, \binits{J.}},
\bauthor{\bsnm{Alfadhl}, \binits{Y.}},
\bauthor{\bsnm{Chen}, \binits{X.}},
\bauthor{\bsnm{Parini}, \binits{C.G.}},
\bauthor{\bsnm{Ying}, \binits{Z.}},
\bauthor{\bsnm{Bolin}, \binits{T.}}:
\batitle{Investigation of on-body bluetooth transmission}.
\bjtitle{IET microwaves, antennas \& propagation}
\bvolume{4}(\bissue{7}),
\bfpage{871}--\blpage{880}
(\byear{2010})
\end{barticle}
\endbibitem

\bibitem{song2013review}
\begin{barticle}
\bauthor{\bsnm{Song}, \binits{Y.}},
\bauthor{\bsnm{Hao}, \binits{Q.}},
\bauthor{\bsnm{Zhang}, \binits{K.}}:
\batitle{Review of the modeling, simulation and implement of intra-body communication}.
\bjtitle{Defence Technology}
\bvolume{9}(\bissue{1}),
\bfpage{10}--\blpage{17}
(\byear{2013})
\end{barticle}
\endbibitem

\bibitem{callejon2015measurement}
\begin{barticle}
\bauthor{\bsnm{Callejon}, \binits{M.A.}},
\bauthor{\bsnm{Reina-Tosina}, \binits{J.}},
\bauthor{\bsnm{Naranjo-Hernandez}, \binits{D.}},
\bauthor{\bsnm{Roa}, \binits{L.M.}}:
\batitle{Measurement issues in galvanic intrabody communication: Influence of experimental setup}.
\bjtitle{IEEE Transactions on Biomedical Engineering}
\bvolume{62}(\bissue{11}),
\bfpage{2724}--\blpage{2732}
(\byear{2015})
\end{barticle}
\endbibitem

\bibitem{maity2018bio}
\begin{barticle}
\bauthor{\bsnm{Maity}, \binits{S.}},
\bauthor{\bsnm{He}, \binits{M.}},
\bauthor{\bsnm{Nath}, \binits{M.}},
\bauthor{\bsnm{Das}, \binits{D.}},
\bauthor{\bsnm{Chatterjee}, \binits{B.}},
\bauthor{\bsnm{Sen}, \binits{S.}}:
\batitle{Bio-physical modeling, characterization, and optimization of electro-quasistatic human body communication}.
\bjtitle{IEEE Transactions on Biomedical Engineering}
\bvolume{66}(\bissue{6}),
\bfpage{1791}--\blpage{1802}
(\byear{2018})
\end{barticle}
\endbibitem

\bibitem{maity2019bodywire}
\begin{barticle}
\bauthor{\bsnm{Maity}, \binits{S.}},
\bauthor{\bsnm{Chatterjee}, \binits{B.}},
\bauthor{\bsnm{Chang}, \binits{G.}},
\bauthor{\bsnm{Sen}, \binits{S.}}:
\batitle{Bodywire: A 6.3-pj/b 30-mb/s- 30-db sir-tolerant broadband interference-robust human body communication transceiver using time domain interference rejection}.
\bjtitle{IEEE Journal of Solid-State Circuits}
\bvolume{54}(\bissue{10}),
\bfpage{2892}--\blpage{2906}
(\byear{2019})
\end{barticle}
\endbibitem

\bibitem{li2021body}
\begin{barticle}
\bauthor{\bsnm{Li}, \binits{J.}},
\bauthor{\bsnm{Dong}, \binits{Y.}},
\bauthor{\bsnm{Park}, \binits{J.H.}},
\bauthor{\bsnm{Yoo}, \binits{J.}}:
\batitle{Body-coupled power transmission and energy harvesting}.
\bjtitle{Nature Electronics}
\bvolume{4}(\bissue{7}),
\bfpage{530}--\blpage{538}
(\byear{2021})
\end{barticle}
\endbibitem

\bibitem{park2019sub}
\begin{barticle}
\bauthor{\bsnm{Park}, \binits{J.}},
\bauthor{\bsnm{Mercier}, \binits{P.P.}}:
\batitle{A sub-10-pj/bit 5-mb/s magnetic human body communication transceiver}.
\bjtitle{IEEE Journal of Solid-State Circuits}
\bvolume{54}(\bissue{11}),
\bfpage{3031}--\blpage{3042}
(\byear{2019})
\end{barticle}
\endbibitem

\bibitem{wen2021channel}
\begin{barticle}
\bauthor{\bsnm{Wen}, \binits{E.}},
\bauthor{\bsnm{Sievenpiper}, \binits{D.F.}},
\bauthor{\bsnm{Mercier}, \binits{P.P.}}:
\batitle{Channel characterization of magnetic human body communication}.
\bjtitle{IEEE Transactions on Biomedical Engineering}
\bvolume{69}(\bissue{2}),
\bfpage{569}--\blpage{579}
(\byear{2021})
\end{barticle}
\endbibitem

\bibitem{nath2022understanding}
\begin{barticle}
\bauthor{\bsnm{Nath}, \binits{M.}},
\bauthor{\bsnm{Ulvog}, \binits{A.K.}},
\bauthor{\bsnm{Weigand}, \binits{S.}},
\bauthor{\bsnm{Sen}, \binits{S.}}:
\batitle{Understanding the role of magnetic and magneto-quasistatic fields in human body communication}.
\bjtitle{IEEE Transactions on Biomedical Engineering}
\bvolume{69}(\bissue{12}),
\bfpage{3635}--\blpage{3644}
(\byear{2022})
\end{barticle}
\endbibitem

\bibitem{zhang2017bioacoustics}
\begin{barticle}
\bauthor{\bsnm{Zhang}, \binits{C.}},
\bauthor{\bsnm{Hersek}, \binits{S.}},
\bauthor{\bsnm{Pu}, \binits{Y.}},
\bauthor{\bsnm{Sun}, \binits{D.}},
\bauthor{\bsnm{Xue}, \binits{Q.}},
\bauthor{\bsnm{Starner}, \binits{T.E.}},
\bauthor{\bsnm{Abowd}, \binits{G.D.}},
\bauthor{\bsnm{Inan}, \binits{O.T.}}:
\batitle{Bioacoustics-based human-body-mediated communication}.
\bjtitle{Computer}
\bvolume{50}(\bissue{2}),
\bfpage{36}--\blpage{46}
(\byear{2017})
\end{barticle}
\endbibitem

\bibitem{bos2018enabling}
\begin{barticle}
\bauthor{\bsnm{Bos}, \binits{T.}},
\bauthor{\bsnm{Jiang}, \binits{W.}},
\bauthor{\bsnm{D’hooge}, \binits{J.}},
\bauthor{\bsnm{Verhelst}, \binits{M.}},
\bauthor{\bsnm{Dehaene}, \binits{W.}}:
\batitle{Enabling ultrasound in-body communication: Fir channel models and qam experiments}.
\bjtitle{IEEE transactions on biomedical circuits and systems}
\bvolume{13}(\bissue{1}),
\bfpage{135}--\blpage{144}
(\byear{2018})
\end{barticle}
\endbibitem

\bibitem{gerguis2021enabling}
\begin{bchapter}
\bauthor{\bsnm{Gerguis}, \binits{J.O.}},
\bauthor{\bsnm{Nath}, \binits{M.}},
\bauthor{\bsnm{Sen}, \binits{S.}}:
\bctitle{Enabling intra-body communication using semi-guided ultrasonic waves through human body tissues}.
In: \bbtitle{Proceedings of Meetings on Acoustics},
vol. \bseriesno{45}
(\byear{2021}).
\bcomment{AIP Publishing}
\end{bchapter}
\endbibitem

\bibitem{modak2022bio}
\begin{barticle}
\bauthor{\bsnm{Modak}, \binits{N.}},
\bauthor{\bsnm{Nath}, \binits{M.}},
\bauthor{\bsnm{Chatterjee}, \binits{B.}},
\bauthor{\bsnm{Maity}, \binits{S.}},
\bauthor{\bsnm{Sen}, \binits{S.}}:
\batitle{Bio-physical modeling of galvanic human body communication in electro-quasistatic regime}.
\bjtitle{IEEE Transactions on Biomedical Engineering}
\bvolume{69}(\bissue{12}),
\bfpage{3717}--\blpage{3727}
(\byear{2022})
\end{barticle}
\endbibitem

\bibitem{nath2019toward}
\begin{barticle}
\bauthor{\bsnm{Nath}, \binits{M.}},
\bauthor{\bsnm{Maity}, \binits{S.}},
\bauthor{\bsnm{Sen}, \binits{S.}}:
\batitle{Toward understanding the return path capacitance in capacitive human body communication}.
\bjtitle{IEEE Transactions on Circuits and Systems II: Express Briefs}
\bvolume{67}(\bissue{10}),
\bfpage{1879}--\blpage{1883}
(\byear{2019})
\end{barticle}
\endbibitem

\bibitem{kibret2014human}
\begin{barticle}
\bauthor{\bsnm{Kibret}, \binits{B.}},
\bauthor{\bsnm{Teshome}, \binits{A.K.}},
\bauthor{\bsnm{Lai}, \binits{D.}}:
\batitle{Human body as antenna and its effect on human body communications}.
\bjtitle{Progress In Electromagnetics Research}
\bvolume{148},
\bfpage{193}--\blpage{207}
(\byear{2014})
\end{barticle}
\endbibitem

\bibitem{li2017evaluation}
\begin{barticle}
\bauthor{\bsnm{Li}, \binits{J.}},
\bauthor{\bsnm{Nie}, \binits{Z.}},
\bauthor{\bsnm{Liu}, \binits{Y.}},
\bauthor{\bsnm{Wang}, \binits{L.}},
\bauthor{\bsnm{Hao}, \binits{Y.}}:
\batitle{Evaluation of propagation characteristics using the human body as an antenna}.
\bjtitle{Sensors}
\bvolume{17}(\bissue{12}),
\bfpage{2878}
(\byear{2017})
\end{barticle}
\endbibitem

\bibitem{park2016channel}
\begin{barticle}
\bauthor{\bsnm{Park}, \binits{J.}},
\bauthor{\bsnm{Garudadri}, \binits{H.}},
\bauthor{\bsnm{Mercier}, \binits{P.P.}}:
\batitle{Channel modeling of miniaturized battery-powered capacitive human body communication systems}.
\bjtitle{IEEE Transactions on Biomedical Engineering}
\bvolume{64}(\bissue{2}),
\bfpage{452}--\blpage{462}
(\byear{2016})
\end{barticle}
\endbibitem

\bibitem{bae2012signal}
\begin{barticle}
\bauthor{\bsnm{Bae}, \binits{J.}},
\bauthor{\bsnm{Cho}, \binits{H.}},
\bauthor{\bsnm{Song}, \binits{K.}},
\bauthor{\bsnm{Lee}, \binits{H.}},
\bauthor{\bsnm{Yoo}, \binits{H.-J.}}:
\batitle{The signal transmission mechanism on the surface of human body for body channel communication}.
\bjtitle{IEEE Transactions on microwave theory and techniques}
\bvolume{60}(\bissue{3}),
\bfpage{582}--\blpage{593}
(\byear{2012})
\end{barticle}
\endbibitem

\bibitem{avlani2020100khz}
\begin{bchapter}
\bauthor{\bsnm{Avlani}, \binits{S.}},
\bauthor{\bsnm{Nath}, \binits{M.}},
\bauthor{\bsnm{Maity}, \binits{S.}},
\bauthor{\bsnm{Sen}, \binits{S.}}:
\bctitle{A 100khz-1ghz termination-dependent human body communication channel measurement using miniaturized wearable devices}.
In: \bbtitle{2020 Design, Automation \& Test in Europe Conference \& Exhibition (DATE)},
pp. \bfpage{650}--\blpage{653}
(\byear{2020}).
\bcomment{IEEE}
\end{bchapter}
\endbibitem

\bibitem{nath2021inter}
\begin{barticle}
\bauthor{\bsnm{Nath}, \binits{M.}},
\bauthor{\bsnm{Maity}, \binits{S.}},
\bauthor{\bsnm{Avlani}, \binits{S.}},
\bauthor{\bsnm{Weigand}, \binits{S.}},
\bauthor{\bsnm{Sen}, \binits{S.}}:
\batitle{Inter-body coupling in electro-quasistatic human body communication: Theory and analysis of security and interference properties}.
\bjtitle{Scientific Reports}
\bvolume{11}(\bissue{1}),
\bfpage{1}--\blpage{15}
(\byear{2021})
\end{barticle}
\endbibitem

\bibitem{sarkar2024wearable}
\begin{bchapter}
\bauthor{\bsnm{Sarkar}, \binits{S.}},
\bauthor{\bsnm{Huang}, \binits{Q.}},
\bauthor{\bsnm{Nath}, \binits{M.}},
\bauthor{\bsnm{Sen}, \binits{S.}}:
\bctitle{Wearable human body communication channel measurements in the body resonance regime}.
In: \bbtitle{2024 IEEE/MTT-S International Microwave Symposium-IMS 2024},
pp. \bfpage{800}--\blpage{803}
(\byear{2024}).
\bcomment{IEEE}
\end{bchapter}
\endbibitem

\bibitem{sarkar2024material}
\begin{bchapter}
\bauthor{\bsnm{Sarkar}, \binits{S.}},
\bauthor{\bsnm{Chowdhury}, \binits{M.R.}},
\bauthor{\bsnm{Huang}, \binits{Q.}},
\bauthor{\bsnm{Sen}, \binits{S.}}:
\bctitle{Material property based analysis of human body communication in body resonance regime}.
In: \bbtitle{2024 IEEE MTT-S International Microwave Biomedical Conference (IMBioC)},
pp. \bfpage{69}--\blpage{71}
(\byear{2024}).
\bcomment{IEEE}
\end{bchapter}
\endbibitem

\bibitem{datta2021advanced}
\begin{botherref}
\oauthor{\bsnm{Datta}, \binits{A.}},
\oauthor{\bsnm{Nath}, \binits{M.}},
\oauthor{\bsnm{Yang}, \binits{D.}},
\oauthor{\bsnm{Sen}, \binits{S.}}:
Advanced biophysical model to capture channel variability for eqs capacitive hbc.
IEEE Transactions on Biomedical Engineering
(2021)
\end{botherref}
\endbibitem

\bibitem{lodi2020periodic}
\begin{barticle}
\bauthor{\bsnm{Lodi}, \binits{M.B.}},
\bauthor{\bsnm{Curreli}, \binits{N.}},
\bauthor{\bsnm{Fanti}, \binits{A.}},
\bauthor{\bsnm{Cuccu}, \binits{C.}},
\bauthor{\bsnm{Pani}, \binits{D.}},
\bauthor{\bsnm{Sanginario}, \binits{A.}},
\bauthor{\bsnm{Spanu}, \binits{A.}},
\bauthor{\bsnm{Ros}, \binits{P.M.}},
\bauthor{\bsnm{Crepaldi}, \binits{M.}},
\bauthor{\bsnm{Demarchi}, \binits{D.}}, \betal:
\batitle{A periodic transmission line model for body channel communication}.
\bjtitle{IEEE Access}
\bvolume{8},
\bfpage{160099}--\blpage{160115}
(\byear{2020})
\end{barticle}
\endbibitem

\bibitem{Gabriel_1996}
\begin{barticle}
\bauthor{\bsnm{{Gabriel et al.}}, \binits{S.}}:
\batitle{The dielectric properties of biological tissues: {II}. measurements in the frequency range 10 hz to 20 {GHz}}.
\bjtitle{Physics in Medicine and Biology}
\bvolume{41}(\bissue{11}),
\bfpage{2251}--\blpage{2269}
(\byear{1996}).
\doiurl{10.1088/0031-9155/41/11/002}
\end{barticle}
\endbibitem

\bibitem{maity2020safety}
\begin{barticle}
\bauthor{\bsnm{Maity}, \binits{S.}},
\bauthor{\bsnm{Nath}, \binits{M.}},
\bauthor{\bsnm{Bhattacharya}, \binits{G.}},
\bauthor{\bsnm{Chatterjee}, \binits{B.}},
\bauthor{\bsnm{Sen}, \binits{S.}}:
\batitle{On the safety of human body communication}.
\bjtitle{IEEE Transactions on Biomedical Engineering}
\bvolume{67}(\bissue{12}),
\bfpage{3392}--\blpage{3402}
(\byear{2020})
\end{barticle}
\endbibitem

\bibitem{ICNIRP2020}
\begin{barticle}
\bauthor{\bsnm{ICNIRP}}:
\batitle{{ICNIRP} guidelines for limiting exposure to electromagnetic fields (100 {kHZ} to 300 {GHZ})}.
\bjtitle{Health Phys.}
\bvolume{74}(\bissue{4}),
\bfpage{483}--\blpage{524}
(\byear{2020})
\end{barticle}
\endbibitem

\bibitem{datta2024S2C}
\begin{botherref}
\oauthor{\bsnm{Datta}, \binits{A.}},
\oauthor{\bsnm{Ding}, \binits{L.}},
\oauthor{\bsnm{Sen}, \binits{S.}}:
Step-to-Charge: mW-scale Power Transfer to On-body Devices for Long Channel ($>$ 1m) with EQS Resonant Human Body Powering.
\url{https://arxiv.org/abs/2408.01927}
\end{botherref}
\endbibitem

\bibitem{ieee1992ieee}
\begin{botherref}
\oauthor{\bsnm{IEEE Standards Coordinating~Committee}, \binits{.}}, et al.:
Ieee standard for safety levels with respect to human exposure to radio frequency electromagnetic fields, 3khz to 300ghz.
IEEE C95. 1-1991
(1992)
\end{botherref}
\endbibitem

\bibitem{cho200960}
\begin{barticle}
\bauthor{\bsnm{Cho}, \binits{N.}},
\bauthor{\bsnm{Yan}, \binits{L.}},
\bauthor{\bsnm{Bae}, \binits{J.}},
\bauthor{\bsnm{Yoo}, \binits{H.-J.}}:
\batitle{A 60 kb/s--10 mb/s adaptive frequency hopping transceiver for interference-resilient body channel communication}.
\bjtitle{IEEE Journal of Solid-State Circuits}
\bvolume{44}(\bissue{3}),
\bfpage{708}--\blpage{717}
(\byear{2009})
\end{barticle}
\endbibitem

\bibitem{sen2016socialhbc}
\begin{bchapter}
\bauthor{\bsnm{Sen}, \binits{S.}}:
\bctitle{Socialhbc: Social networking and secure authentication using interference-robust human body communication}.
In: \bbtitle{Proceedings of the 2016 International Symposium on Low Power Electronics and Design},
pp. \bfpage{34}--\blpage{39}
(\byear{2016})
\end{bchapter}
\endbibitem

\end{thebibliography}

\section*{Acknowledgments:}
This work was supported by Quasistatics, Inc. dba Ixana –Grant 40003567. The authors thank Lingke Ding and David Yang, PhD students at Sparclab for their valuable input during the experimental process.
\section*{Authors' contributions:}
S. Sarkar, M. Nath, S. Sen conceived the idea. M.Nath was at Purdue University during his contribution to this work and also provided useful suggestions on the theory development and experiments thereafter. S. Sarkar and S. Sen conducted the theoretical analysis. S. Sarkar and Q. Huang conducted numerical simulations. S. Sarkar, S. Antal and Q. Huang performed the experiments. All the authors analyzed the results and reviewed the manuscript.\newline 
\section*{Competing interests:} 
The authors declare that S. Sen have a financial interest in Quasistatics, Inc.\newline
\section*{Additional information:}
The supplementary information is available at
\url{https://github.com/SparcLab/BodyResonanceHBC}
\\
Correspondence and requests for materials should be addressed to Shreyas Sen (shreyas@purdue.edu) or Samyadip Sarkar.


\end{document}